\documentclass[prb,twocolumn,floatfix,showpacs,groupedaddress]{revtex4}
\usepackage{graphicx,amsfonts,amssymb,amsmath,hyperref,nicefrac,lipsum, verbatim}

\usepackage[version=3]{mhchem}

\usepackage{siunitx}  
\usepackage[x11names]{xcolor}
\newif\ifhyper
\hypertrue
\ifhyper
\hypersetup{
   citecolor = {green!60!black},
   colorlinks = {true}, 
   urlcolor = {blue} 
}
\fi

\newcommand{\beq}{\begin{equation}}
\newcommand{\eeq}{\end{equation}}
\newcommand{\beqa}{\begin{eqnarray}}
\newcommand{\eeqa}{\end{eqnarray}}
\newcommand{\ket} [1] {\vert #1 \rangle}

\newcommand{\braket}[2]{\langle #1 | #2 \rangle}

\def\ket#1{\vert#1\rangle}

\def\Longarrow{\protect\@lra}
\def\@lra{\relbar\joinrel\relbar\joinrel\relbar\joinrel%
          \relbar\joinrel\rightarrow}

\def\corr#1{\langle#1\rangle}

\begin{document}

\title{Tunable edge states and their robustness towards disorder}

\author{M.~Malki}
\email{maik.malki@tu-dortmund.de}
\affiliation{Lehrstuhl f\"ur Theoretische Physik 1, TU Dortmund, Germany}

\author{G.\ S.~Uhrig}
\affiliation{Lehrstuhl f\"ur Theoretische Physik 1, TU Dortmund, Germany}
\date{\rm\today}

\begin{abstract}
The interest in the properties of edge states in Chern insulators and in 
$\mathbb{Z}_2$  topological insulators has increased rapidly in recent years. 
We present calculations on how to influence the transport properties of chiral and helical edge states by modifying the edges in the Haldane and in the 
Kane-Mele model.  The Fermi velocity of the chiral edge states becomes 
direction dependent as does the spin-dependent Fermi velocity of 
the helical edge states. 
Additionally, we explicitly investigate the robustness of edge states 
against local disorder. The edge states can be reconstructed in the 
Brillouin zone in the presence of disorder. The
influence of the width and of the length of the system is studied as well as 
the dependence of the edge states on the strength of the disorder.
\end{abstract}


\maketitle

\section{Introduction}
\label{sec:introduction}

\subsection{General context}

Since the discovery of the integer and the fractional quantum Hall effect 
\cite{klitz80,tsui82}
topological phenomena have become an important field of research in condensed matter physics. The edge states \cite{hatsu93} occurring  in the quantum Hall 
effect are localized exponentially at the boundaries of the sample.  They
are protected by the topological properties of the band structure in the bulk. 
As shown by Thouless et al.\ \cite{thoul82} the number of edge states 
corresponds to the Chern number $\nu$ of the filled electronic bands
and implies the famous quantized Hall conductivity $\sigma_{xy} = \nu e^2/ h$. 
The description of the quantum Hall effect by topological invariants
\cite{avron83,niu85,kohmo85,uhrig91} is based on the  Berry phase \cite{berry84}.

In order to mimic the integer quantum Hall effect in a lattice model 
\emph{without} external magnetic field Haldane has proposed the first Chern insulator \cite{halda88b}. In addition to nearest-neighbor hopping on a honeycomb lattice, the proposed Haldane model comprises
a staggered magnetic flux which induces complex next-nearest-neighbor hopping
elements while the translational symmetry is preserved. 
Averaged over a unit cell of the lattice
the magnetic flux vanishes. However, for certain values of the phases
a finite magnetic field cannot be distinguished from a vanishing average magnetic
field because phases of multiples of $2\pi$ cannot be distinguished from
vanishing phases \cite{redde16}.

In order to realize a Chern insulator the time reversal symmetry (TRS) must be broken.  Nontrivial Chern numbers imply chiral edge states also in the absence of an external magnetic field extending the concept of the usual quantum Hall effect. In the context of topologically protected edge states the term `chiral' means that the electrons  only propagate in one direction along one edge.
If no magnetic field is involved this effect is called  the \emph{anomalous}
quantum Hall effect \cite{weng15,liu16,ren16}.

The Kane-Mele model \cite{kane05a,kane05b,hasan10} represents
a crucial extension of the Haldane model including the spin degree of freedom.
This renders the preservation of the TRS possible because one spin species 
realizes the time-reversed replica of the other. The
Kane-Mele model was suggested to describe the effect of 
spin-orbit interaction on the electronic band structure of graphene in the 
low-energy regime, but it turned out that the spin-orbit interaction in 
graphene is too weak to produce noticeable effects.
Nevertheless, the Kane-Mele model provides fascinating theoretical insights.

Due to the preservation of the TRS in the Kane-Mele model it cannot display 
a net charge current at the edges. Instead, a net spin current is possible.
This phenomenon is referred to as the quantum spin Hall effect (QSHE) which 
can be attributed to the topological $\mathbb{Z}_2$ 
invariant \cite{kane05b, fu06, berne13} implying helical edge states \cite{wu06}. 
These topologically protected edge states are called `helical' because they have a spin filtering  property, i.e., 
the \smash{spins $\uparrow$} propagate in one direction while 
the spins $\downarrow$ propagate in the opposite direction along the same edge. 
As a result, the QSHE implies a quantization of the spin Hall conductivity.
Basically, materials displaying the QSHE and  characterized
 by the $\mathbb{Z}_2$ topological invariant are referred to as 
 $\mathbb{Z}_2$ topological insulators \cite{hasan10, qi11}.

Since the QSHE is too weak in graphene to be measurable, 
Bernevig, Hughes, and Zhang proposed a model \cite{berne06} 
for the QSH phase in \ce{HgTe} quantum wells where the
spin-orbit coupling is much stronger. Soon after the theoretical proposal the
 QSH phase has been  observed experimentally in a 
2D \ce{HgTe}/\ce{CdTe} quantum well 
\cite{konig07,roth09}. Another experimental observation  of the QSHE  was realized
in \ce{InAs}/\ce{GaSb} quantum wells \cite{knez11,du15}. The discovered QSHE is
only measured at low temperatures up to \SI{40}{K}. Theoretical 
calculations \cite{liu11} predict a possible realization of the 
QSHE in germanium with a large energy gap corresponding to \SI{277}{K}. 
The calculated energy gap of the low-buckled honeycomb structure of germanium results from the stronger spin-orbit coupling so that this
system is a candidate for detecting the QSHE at higher temperatures.

Historically, the QSHE  was measured first in 2D topological insulators.
A Chern insulator was considered unlikely to be realized.
But very recently, progress has been achieved towards 2D Chern insulators.
 The first observation of the anomalous quantum Hall effect was made in thin 
ferromagnetic Chern insulators  \cite{chang13,kou14,chang15}. It could be observed
up to temperatures of a few Kelvin. Theoretical proposals indicate that Chern 
insulators near room temperature are possible 
in thin ferromagnetic Chern insulators \cite{wu14} 
or in superlattices of gold atoms on single-vacancy graphene \cite{han15,krash11}.

An alternative realization has been achieved using ultracold fermionic \ce{^{40}K}
 atoms in a periodically  modulated optical lattice \cite{jotzu14}. This method 
could implement the  Haldane model in an experimental setup. 
A particular asset of this route is the tunability of the physical properties.

\subsection{Present objectives}

Due to their topological protection, edge states may carry currents 
without dissipation and they are protected against disorder to some extent, 
see below. This robustness makes them attractive to applications.
With this long-term goal in mind, we set out to study
 the influence of controllable external parameters on the transport behavior 
of topological edge state as well as to study 
the effect of noncontrollable features such as
disorder. We do not focus on the DC conductivity as has been done before, see 
for instance Refs.\ \onlinecite{dyke16, qiao16}. We choose the Fermi velocity 
$v_\text{F}$ as the measurable quantity of interest 
in order to gain understanding which is complementary to the existing literature.

The Fermi velocity is a key quantity in transport behavior representing  the
group velocity of a transmitted charge or spin signal. Thus, we aim at tuning
the Fermi velocity which quantifies how fast a signal is transmitted. A previous observation in the Kagome lattice
 \cite{redde16} revealed that the Fermi velocity depends on the 
chosen shape of the edge. 
Further investigations in the Haldane model \cite{uhrig16}
showed that by decorating one edge of the  honeycomb lattice
the Fermi velocity can be influenced strongly.
We extend this observation by considering decorating both edges.

Next, we transfer the idea of decoration to the Kane-Mele model, i.e., 
to helical edge states. 
An explicit Rashba coupling \cite{bychk84} and its effect on the Fermi 
velocity is also studied. The decoration of the edges of the Kane-Mele 
model leads to a tunable spin-dependent Fermi velocity which suggests the applicability of tunable transmission speeds in spintronics 
\cite{wolf01,zutic04}. 

Finally, we study the influence of local disorder on the edge states. 
Since edge modes are due to nontrivial topological invariants 
it is assumed that they are protected against  disorder. However, various experiments show that the signatures of topological phases are much more prominent
in  high-purity samples \cite{hasan10} than in samples of lower quality.
Thus, we intend to investigate the influence of disorder by explicit 
calculations. For simplicity,  we study the robustness of the 
chiral edge states in the Haldane model on the honeycomb lattice.

Local disorder breaks the translational invariance. By calculating the 
modulus squared of the overlap of the eigenwave functions of the disordered 
system with the eigenedge modes in 
the clean system we define a transition probability. 
The maximization of this quantity is used to reconstruct the 
momenta of the edge states. The dependence of the
the transition probability on the width, the length of the system as well as 
on the strength of the local disorder is examined. 
We find that the disorder may not exceed certain
thresholds in order to preserve the characteristic transport 
properties of the edge modes.

\section{Decoration of the Haldane model}
\label{sec:haldane}

For the sake of completeness, we recap results for decorated 
edges in the model without spin \cite{uhrig16}.
The results are important for the comparison
with the results in the modified and extended models with spin.
Moreover, they serve as reference for the disordered Haldane model which we investigate in Sec.\ \ref{sec:robustness}.

\begin{figure}
	\centering
		\includegraphics[width=1.0\columnwidth]{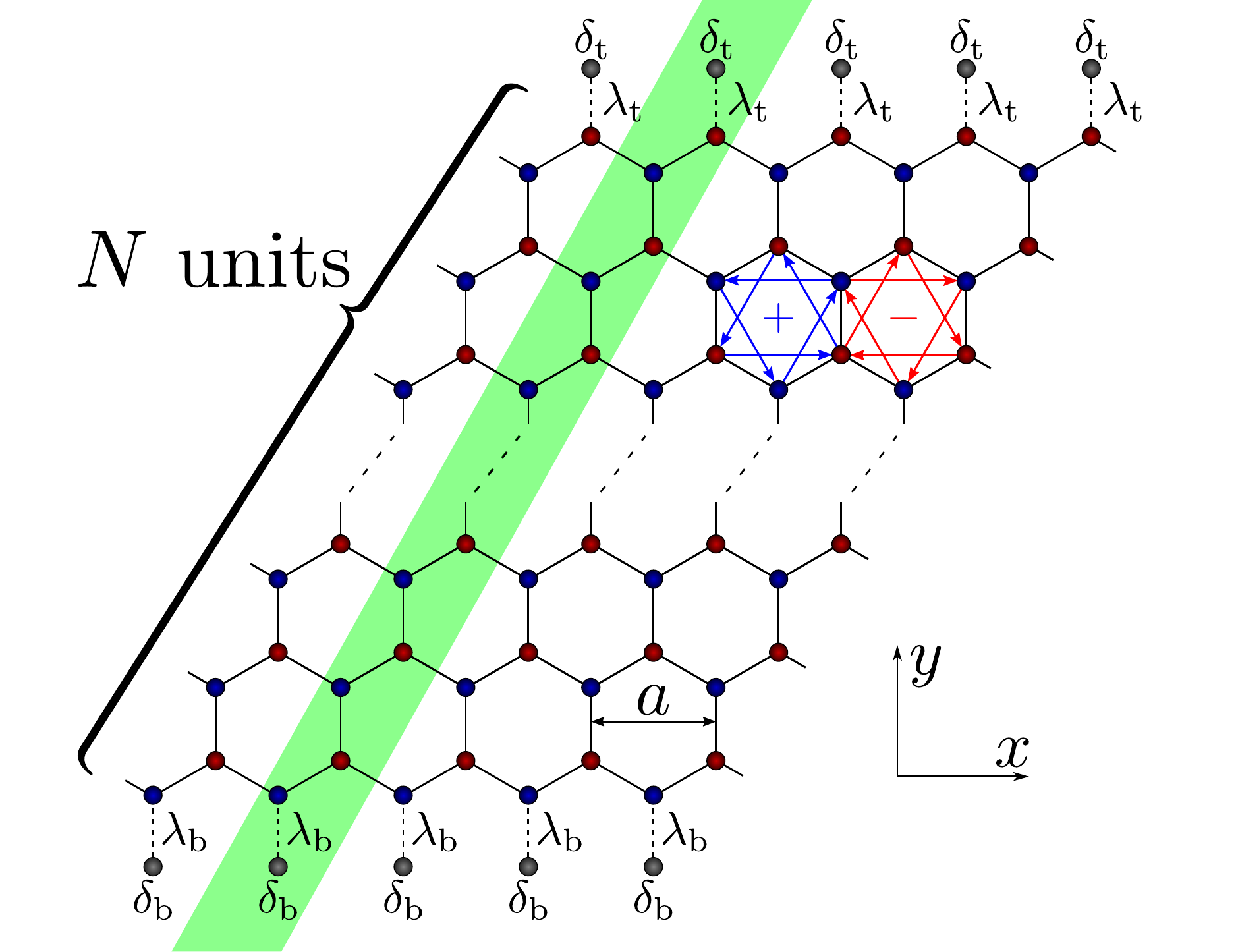}
	\caption{(Color online) Sketch of a strip in the honeycomb lattice with NN hopping (black bonds). The green (light shaded) area displays a unit cell in the 
	$x$ direction which consists of $2N+2$ sites including the decorating sites.
	The honeycomb with blue arrows illustrates the complex hopping elements to NNN 
	sites with phase $\phi$ while the honeycomb with red arrows illustrates
	the hopping elements with phase $-\phi$, see Eq.\ \eqref{eq:sign}.
  The top and the bottom edge are decorated by additional sites which are 
	coupled weakly ($\lambda_{\gamma} < t$) to the bulk sites. 
	These decorating sites are subject to a
	local potential $\delta_{\gamma}$. The index $\gamma$ 
	takes the value $\text{t}$ (top) or $\text{b}$ (bottom). 
	The lattice constant $a$ is set to unity.}
	\label{fig:strip}
\end{figure}

The complete Hamiltonian of the model can be divided into two contributions
\begin{subequations}
\begin{equation}
	\mathcal{H} = \mathcal{H}_{\mathrm{strip}} + \mathcal{H}_{\mathrm{decor}}
\end{equation}
with
\begin{eqnarray}
	\mathcal{H}_{\mathrm{strip}} &=& t \sum_{\corr{ i, j }} c_i^{\dagger} c_j^{\phantom{\dagger}} 
	+ t_2\sum_{\corr{\corr{i, j}}} \mathrm{e}^{\mathrm{i} \nu_{ij} \phi} 
	c_i^{\dagger} c_j^{\phantom{\dagger}} 
	\label{eq:haldane}\\
	\mathcal{H}_{\mathrm{decor}} &=& \sum_{i \gamma} \left[ \lambda_{\gamma}
	\left(c^{\dagger}_{d(i)} c_i^{\phantom{\dagger}} + c^{\dagger}_i c_{d(i)}^{\phantom{\dagger}} \right) + \delta_{\gamma} 
	c^{\dagger}_{d(i)} c_{d(i)}^{\phantom{\dagger}} \right] . \qquad
\end{eqnarray}
\end{subequations}
The corresponding honeycomb lattice with decorated edges is shown in 
Fig.\ \ref{fig:strip}. 
The Hamiltonian $\mathcal{H}_{\mathrm{strip}}$ comprises a real hopping
element between  nearest neighbor (NN) sites and a complex hopping element
between  next-nearest neighbor (NNN) sites. 
The symbol $\corr{i, j}$ denotes a pair of NN sites 
while $\corr{\corr{i, j}}$ denotes a pair of NNN sites. 
The hopping parameter $t$ is real and serves as energy unit henceforth. 
The lattice constant $a$ is set to unity. The 
complex hopping element is given by the combination of the 
positive real parameter $t_2$  and a \smash{phase $\phi$}. 
The inclusion of the nonvanishing phase breaks the TRS as is 
necessary for obtaining nontrivial Chern numbers. 

The sign of the phase is determined by
\begin{equation}
	\nu_{ij} = \mathop{\mathrm{sgn}}(\hat{d}_1(ij) \times \hat{d}_2(ij))_z 
	= \pm 1
	\label{eq:sign}
\end{equation} 
where one reaches site $j$ from site $i$ by a NN hop to site $i'$
and a second NN hop from $i'$ to $j$. Then,
 $\hat{d}_1(ij)$ stands for the unit vector 
from $i$ to $i'$ and $\hat{d}_2(ij)$ for the unit vector from $i'$ to $j$.
The complex hopping elements with \smash{phase $\phi$} 
are shown in blue (dark gray)  while the hopping elements with phase 
$-\phi$ are shown in red (light gray) in Fig.\ \ref{fig:strip}. 

The Hamilton operator of the decorating sites 
$\mathcal{H}_{\mathrm{decor}}$ consists of two parts. 
One part describes the additional sites at the top whereas the other part describes the bottom sites ($\gamma \in \{\text{t}, \text{b}\}$).
If the outermost sites of the undecorated honeycomb lattice are 
denoted by $j$ the adjacent decorating sites are labeled $d(j)$. 
The hopping elements between the outermost sites and the attached decorating sites 
are modified by the factor $\lambda_{\gamma}$. Generically, we consider an
attenuation so that $0 \leq \lambda_{\gamma} \leq t$ holds. 
The on-site energy of the decorating sites are denoted by $\delta_{\gamma}$.
It can be thought to be generated by a gate voltage which changes the 
electric potential of the decorating sites \cite{uhrig16}.

The  phase diagram of the Haldane model on a bulk honeycomb lattice without boundaries can be found for instance in Refs.\ 
\onlinecite{halda88b, berne13}. 
Calculating the dispersion on a finite strip, see Fig.\ \ref{fig:strip}, 
of the system provides the chiral edge states. 
Coupling the decorating sites to the honeycomb strip does not alter the 
topological characteristics of the system. The phase $\phi$ is set to $\pi/2$ in
 order to maximize the gap. To create rather flat energy bands we set 
\smash{$t_2 = 0.2 t$} as in \smash{Ref.\ \onlinecite{uhrig16}.} 
The Fermi level is set to $\varepsilon_{\mathrm{F}} = 0$.

To illustrate the impact of the modification we calculate the dispersion 
of both edge modes and compare it to the dispersion in the 
undecorated Haldane model. In the following, 
we investigate a strip of finite height in the $y$ direction whereas the strip is infinitely extended in the $x$ direction, see \smash{Fig.\ \ref{fig:strip}}. Due to the translational symmetry  in the $x$ direction, the wave number $k_x$ 
represents a good quantum number. At fixed $k_x$, one obtains a  
$(2 N + 2) \times (2 N + 2)$ one-particle matrix which can be diagonalized numerically. The dispersive modes within the gap of the bulk
Haldane model are the edge modes. Due to their exponential localization 
at the edges their dispersion converges quickly upon increasing the 
\smash{number $N$} of units  in the strip. 
The calculations in this work are based on strips with $N = 60$ units
which turns out to be sufficiently wide.

\begin{figure}
	\centering
	\includegraphics[width=\columnwidth]{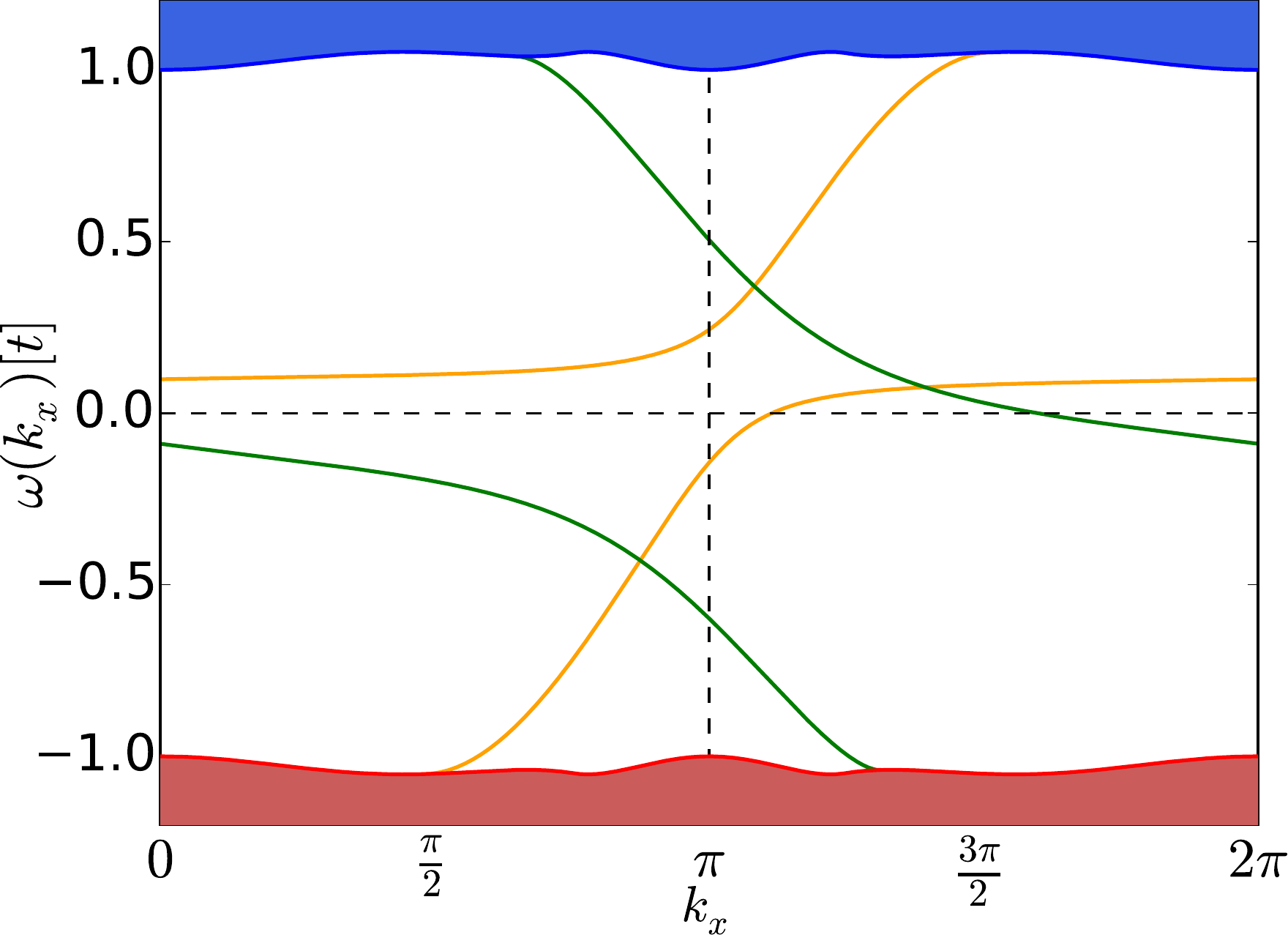}
	\caption{(Color online) (a) Dispersion of the edge states with 
	\smash{$t_2 = 0.2 t$}, 
	$\phi = \pi/2$, $\lambda_\mathrm{t} = 0.2 t$, $\delta_\mathrm{t} = 0.1 t$, $\lambda_\mathrm{b} = 0.6 t$, and 
	\smash{$\delta_\mathrm{b} = -0.1 t$}. The right-moving edge state marked in orange (light gray) is located at the top edge while the left-moving edge state marked in green (dark gray) is located at the
	the bottom edge. The filled areas indicate the continua of the bulk states. The Fermi velocity $v_{\mathrm{F}} = \partial \omega / \partial k_x |_{\epsilon_{\mathrm{F}}}$ of both edges modes are independent of each other. }
	\label{fig:tune_haldane}
\end{figure}

An example of a dispersion with different parameters for both edges is shown 
in Fig.\ \ref{fig:tune_haldane}. The filled areas represent the continua 
stemming from the  modes for all possible values of $k_y$. 
Our main focus lies on the investigation of the edge modes
of which the energies are between the lower band edge
of the upper continuum (blue, darker shading) and the 
upper band edge of the lower continuum (red, lighter shading).

Upon coupling the decorating sites to the honeycomb strip, 
i.e., $\lambda_{\gamma} \neq 0$, the dispersions of the edge modes 
display an `avoided crossing' (or `level repulsion') due to the 
hybridization \smash{with the} local modes from the decorating sites.
In the case of small values of $\lambda$, see right mover in 
Fig.\ \ref{fig:tune_haldane}, the edge states have a rather flat band.
Increasing $\lambda$ leads to a stronger repulsion between the edge
 modes near the zone boundary $k_x = \pi$ so that the dispersion acquires stronger momentum dependence, see left mover in Fig.\ \ref{fig:tune_haldane}. 

Besides the coupling $\lambda_{\gamma}$, the decorating sites can 
be influenced by the local potentials $\delta_{\gamma}$. 
Increasing the local energy of the decorating sites counteracts the 
hybridization because the tendency of an electron
to visit the decorating sites is decreased if these sites differ
in energy from the bulk sites. In this way, the 
decorating sites can be smoothly switched off. Then, 
the Fermi velocity converges to 
the Fermi velocity $v_{\mathrm{F}0}$ without decoration. 

The dependence of the Fermi velocity on the parameters of the decorated model
has been studied quantitatively for a single decorated edge \cite{uhrig16}. 
To prove the independence of the chiral edge modes explicitly we calculated
the Fermi velocity of both edges while tuning parameters of only one edge.
The Fermi velocity of the unaltered edge remains unaffected to the tenth digital.
We stress that the relative coupling $\lambda_{\gamma}$ and the local potential $\delta_{\gamma}$ provide controllable parameters to tune the Fermi velocity 
of the edge mode independent from the other edge mode. Furthermore, different 
decorations at the top and at the bottom edge enable us to realize different 
Fermi velocities $v_{\mathrm{F},\gamma}$ so that 
the velocities become direction-sensitive.

\section{Decoration of the Kane-Mele model}
\label{sec:kane_mele}

Here, we investigate the impact of decorated edges on the helical edge states of the Kane-Mele model which includes the spin degree of freedom in such a fashion that it preserves the TRS. The  Hamiltonian reads
\begin{subequations}
\begin{equation}
	\mathcal{H} = \mathcal{H}_{\mathrm{strip}} + \mathcal{H}_{\mathrm{decor}}
\label{eq:kane_mele}
\end{equation}
with
\begin{eqnarray}
	\mathcal{H}_{\mathrm{strip}} &=& t \sum_{\corr{ i, j } \alpha} c_{i\alpha}^{\dagger} 
	c_{j\alpha}^{\phantom{\dagger}} + \mathrm{i} t_2 \sum_{\corr{\corr{i, j}}\alpha \beta} \nu_{ij} 
	c_{i\alpha}^{\dagger} \sigma^z_{\alpha \beta} c_{j\beta}^{\phantom{\dagger}} 
	\nonumber \\
	&& + \mathrm{i} t_\mathrm{r} \sum_{\corr{i, j}\alpha \beta} c_{i\alpha}^{\dagger} 
	(\sigma_{\alpha\beta} \times \hat{d}_{ij})_z c_{j\beta}^{\phantom{\dagger}}
	\label{eq:kane_mele1}
\end{eqnarray}
and
\begin{eqnarray}
	\mathcal{H}_{\mathrm{decor}} &=& \sum_{i \gamma \alpha} \left[ \lambda_{\gamma} 
	\left(c^{\dagger}_{d(i)\alpha} c_{i\alpha}^{\phantom{\dagger}} + c^{\dagger}_{i\alpha} c_{d(i)\alpha}^{\phantom{\dagger}} \right) 
	\right. \nonumber\\ 
	\phantom{\mathcal{H}_{\mathrm{decor}}} && \phantom{\sum_{i \gamma \alpha}} \, \, \left. + 
	\delta_{\gamma} c^{\dagger}_{d(i)\alpha} c_{d(i)\alpha}^{\phantom{\dagger}} \right] 
		\label{eq:kane_deco}
\end{eqnarray}
\end{subequations}
on the honeycomb lattice similar to
the decorated Haldane model from the previous section,
see Fig.\ \ref{fig:strip}. In the Kane-Mele model, each site can host two electrons
with spin quantum number denoted by $\alpha, \beta \in 
\left\lbrace \uparrow, \downarrow \right\rbrace$. 
The Hamilton operator of the strip contains three contributions. The first term describes the usual tight-binding hopping $t$ between NN sites. 
As before the hopping parameter $t$ is real and used as the energy unit. 

Kane and Mele \cite{kane05a} argued that the second hopping \smash{term $\propto t_2$}
is induced by spin-orbit interaction. The hopping parameter $t_2$ is real
and the sign depends on the NNN sites $i$ and $j$ as given by $\nu_{ij}$
defined in \eqref{eq:sign}. The NNN term is closely related
 to the NNN hopping in the Haldane model. 
Considering each spin species separately, the corresponding Hamiltonian with NN and NNN hopping violates the TRS.  It equals the
 Haldane Hamiltonian at $\phi = \pm \pi/2$ for either $S^z=+1/2$
or $S^z=-1/2$. The Kane-Mele model comprises two decoupled 
Haldane models with opposite phases. Since the time reversal 
transformation \smash{$T = \exp(- \mathrm{i} \pi S^y) K$}
 maps one onto the other their combination preserves the TRS \cite{berne13}. 

The last term in $\mathcal{H}_{\mathrm{strip}}$ proportional to $t_\mathrm{r}$ 
describes a Rashba term \cite{kane05a,bychk84} which can also result from 
spin-orbit coupling in the presence of a perpendicular electric 
field or a certain interaction with a substrate. 
The hopping element \smash{parameter $t_\mathrm{r}$} of the Rashba term is real.
The Rashba term violates the conservation of the total $S^z$ component 
so that the two Haldane models for $S^z=\pm1/2$ are coupled for 
$t_\mathrm{r}\neq 0$.

The Hamiltonian of the decorating sites at the edge is chosen
to be spin independent for simplicity, 
similar to the decoration of the Haldane model. 
So the notation will be the same except 
that an additional index is used to denote the spin.

The topological phases of the Kane-Mele model are classified by a $\mathbb{Z}_2$ 
invariant. The phase diagram of the bulk Kane-Mele model including the 
Rashba coupling  is known \cite{kane05b,berne13}. 
We detect the presence of helical edge states by 
calculating the dispersion on a strip of finite width as before. 

First, we set the Rashba coupling to zero so that our results can be directly linked to the results for the decorated Haldane model. For 
$\lambda_{\gamma} = \delta_{\gamma} = 0$, the original Kane-Mele model 
on a strip is retrieved. The corresponding Hamiltonian consists of 
two decoupled Haldane Hamiltonian each of which displays its own chiral edge states. The chiral edge states of the  spin $\uparrow$ part move in 
opposite direction to the chiral edge states of the spin $\downarrow$ part 
because the phase of their NNN hopping element is opposite. The two chiral edge 
states with opposite spins constitute a pair of counterpropagating 
edge modes at each edge.  As shown in the previous section, the 
top edge can be modified independently of the bottom edge. 
This also holds true for the Kane-Mele model. 
Therefore, we only consider the decoration of the top edge 
 in the following for brevity.

\begin{figure}
	\centering
		\includegraphics[width=\columnwidth]{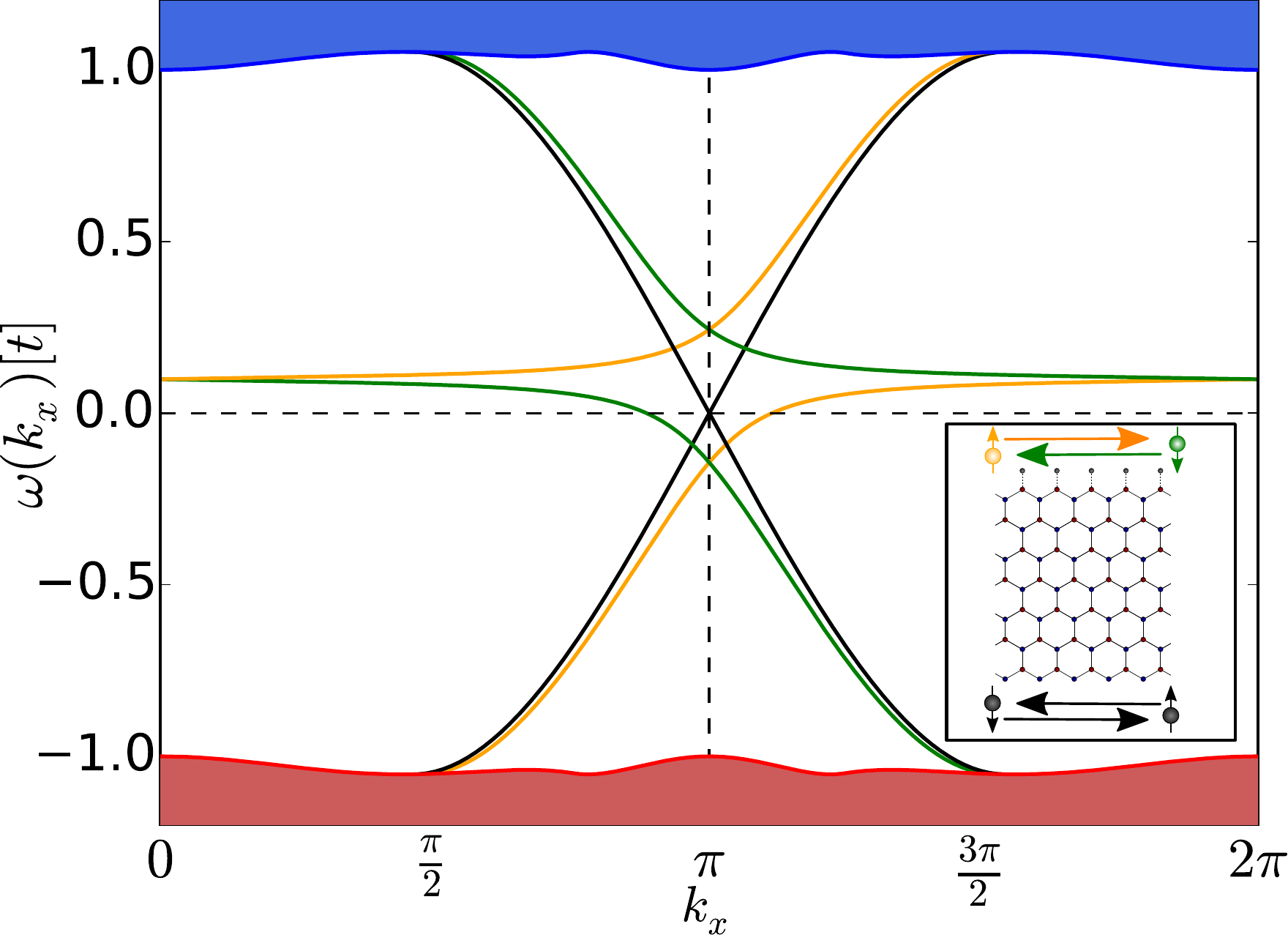}
	\caption{(Color online) Dispersion of the edge states with \smash{$t_2 = 0.2 t$}, 
	$\lambda_\mathrm{t} = 0.2 t$ and $\delta_\mathrm{t} = 0.1 t$. The filled areas
	indicate the continua of the bulk states. The edge states located at the top 
	edge are shown in color (gray).	The spin $\uparrow$ mode propagating to the 
		right is marked in orange (light gray) while 
	the spin 	$\downarrow$ mode counterpropagating to the left is marked in green 
	(dark gray). 
	The dispersion of the edge states at the lower boundary are displayed in black. 
	The schematic sketch in the inset clarifies the assignments.}
	\label{fig:kanemele}
\end{figure}

In the Kane-Mele model, the Fermi velocities of the edge modes 
are  spin-dependent. Except for this difference, one can carry over the 
basic considerations  that we developed for the decorated Haldane model. 
Figure \ref{fig:kanemele} illustrates this point.
The helical edge states of the bottom edge are  
the same edge states as in an undecorated Kane-Mele strip 
because the bottom edge is undecorated. 
The dispersion of the modes at the top edge display
the effect of the `avoided crossing' combined with a certain shift of
the dispersion due to the local potential. This is in line with the results
for the Haldane model.

\begin{figure}
	\centering
		\includegraphics[width=\columnwidth]{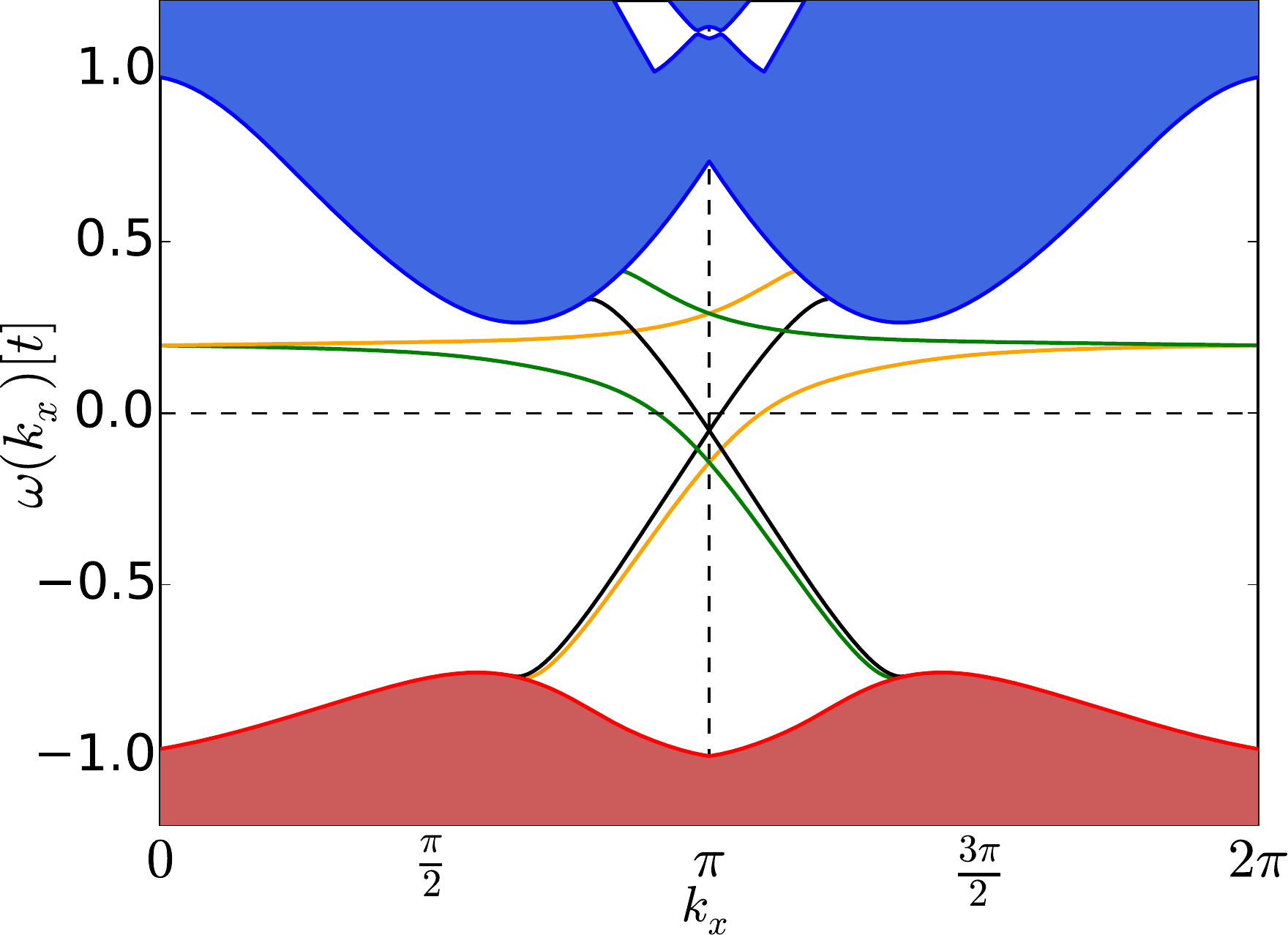}
	\caption{(Color online) Dispersion of the edge states at finite Rashba coupling
	at	\smash{$t_2 = 0.2 t$}, $t_\mathrm{r} = 0.8 t$, $\lambda_\mathrm{t} = 0.2 t$
	and $\delta_\mathrm{t} = 0.2 t$. The 
	filled areas indicate the continua of the bulk states. 
	The edge states located at the top edge are shown in color (gray).
	The spin $\uparrow$ mode propagating to the right is marked in orange 
	(light gray) while the spin 	$\downarrow$ mode 
	counterpropagating to the left is marked in green (dark gray). 
	The dispersions of the edge states at the lower boundary are displayed in black. 
	}
	\label{fig:kanemele_ras}
\end{figure}

Due to TRS, the dispersions display two mirror planes at the  momenta 
invariant under time reversal: $k_x = 0$ or $k_x = \pi$. 
This property is based on Kramer's theorem \cite{krame30}. 
Kramer's theorem predicts crossing points of the counterpropagating 
edge states at the time reversal invariant momenta. The two crossing modes represent
the famous Kramer's pairs. Their degeneracy is robust against 
time reversal symmetric perturbations. 
The level repulsion of the `avoided crossing' leads to a Kramer doublet 
located at $k_x = 0$. The number of Kramer's doublets at one edge must be odd
in the topologically nontrivial phase because it is related to the 
$\mathbb{Z}_2$ topological invariant \cite{hasan10}. 

\begin{figure}
	\centering
		\includegraphics[width=1.0\columnwidth]{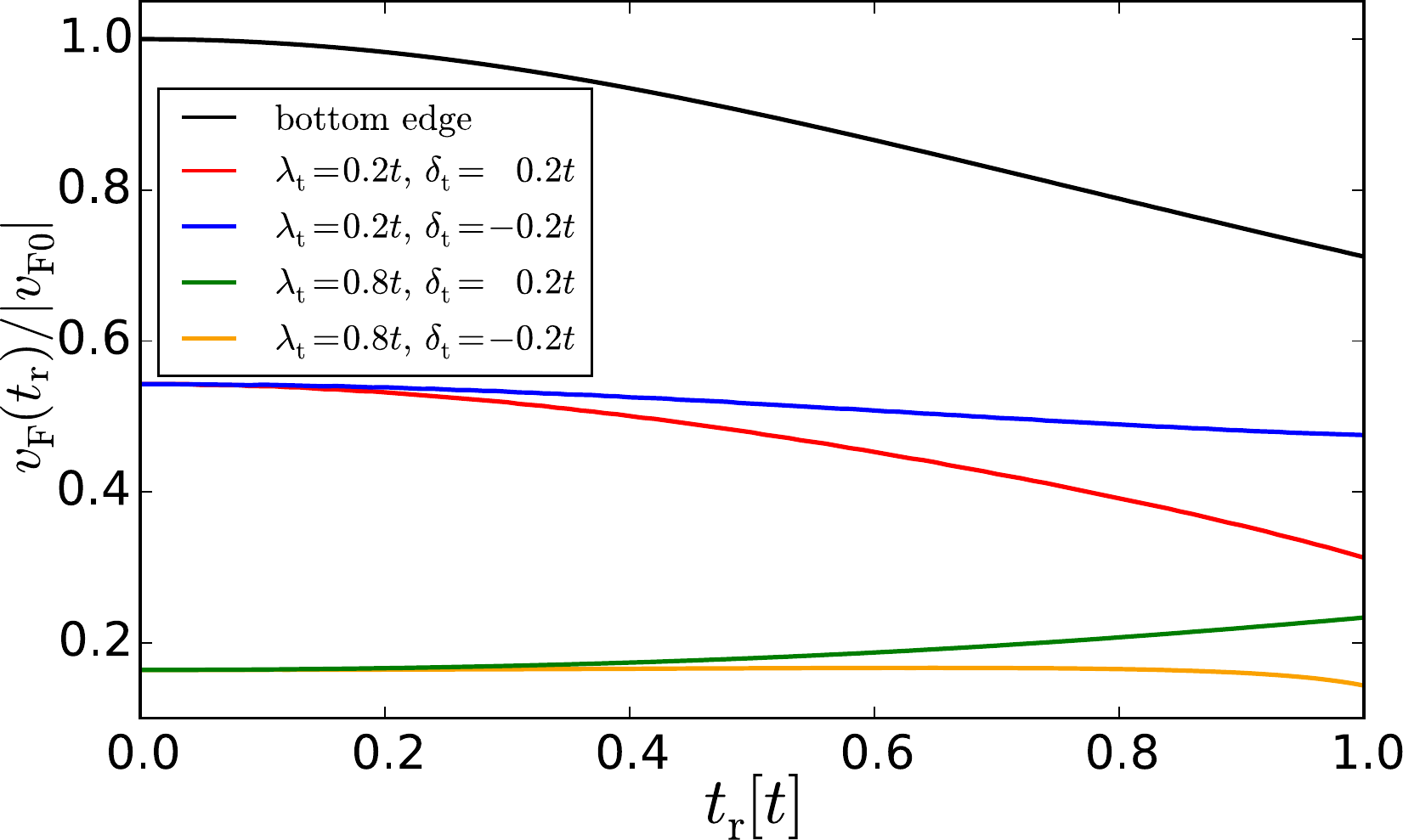}
	\caption{(Color online) Fermi velocity $v_{\mathrm{F}} = 
		\partial \omega / \partial k_x |_{\varepsilon_{\mathrm{F}}}$ of the
		right-moving edge modes relative to the original $v_{\mathrm{F}0}$
		without decorated edges vs.\ the 	Rashba coupling $t_\mathrm{r}$ at 
		$t_2 = 0.2 t$ for various decorations of the top edges 
		and undecorated bottom edges.}
	\label{fig:rashba}
\end{figure}

\begin{figure*}
	\centering
	\includegraphics[width=\columnwidth]{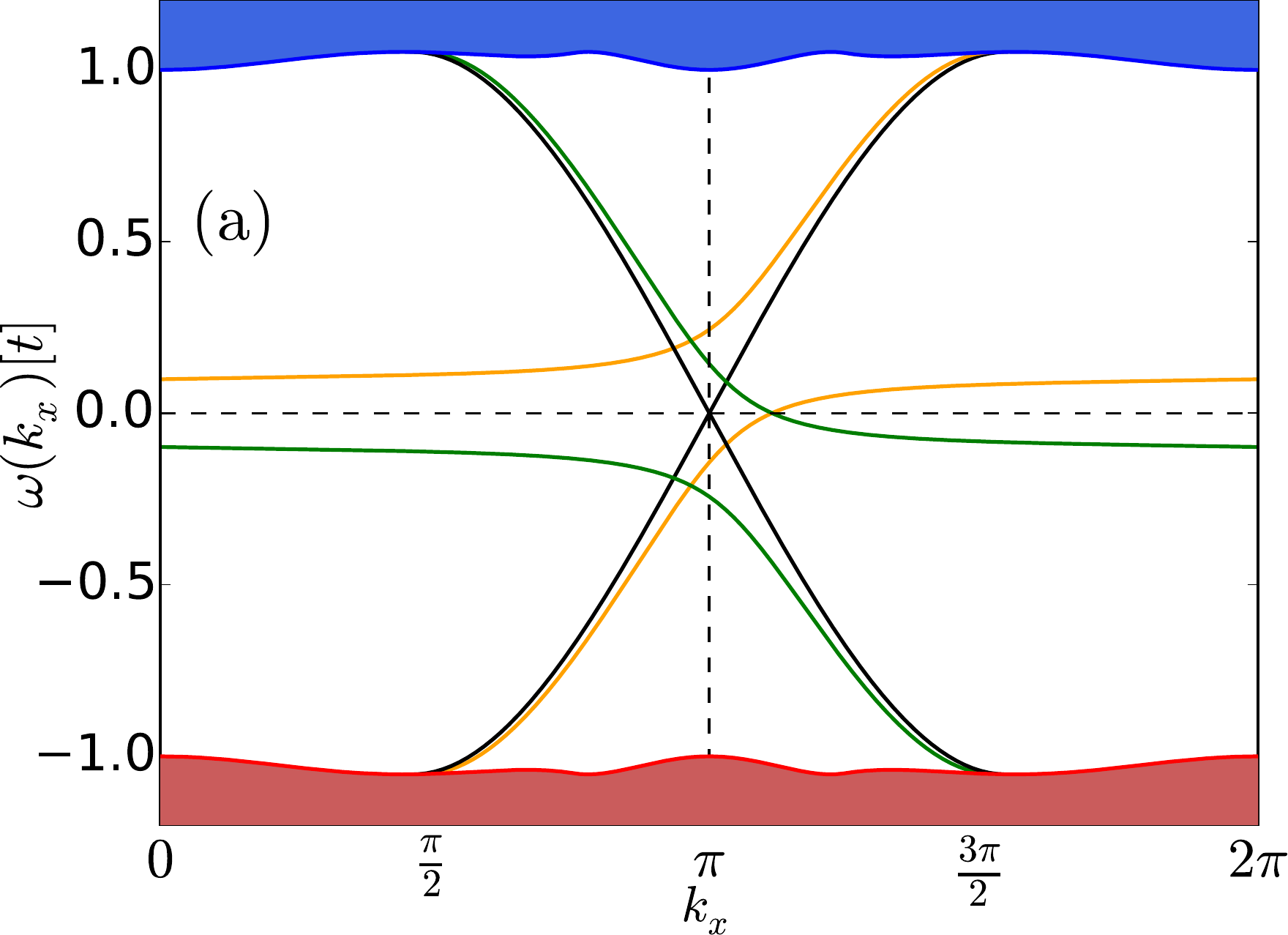}
	\includegraphics[width=\columnwidth]{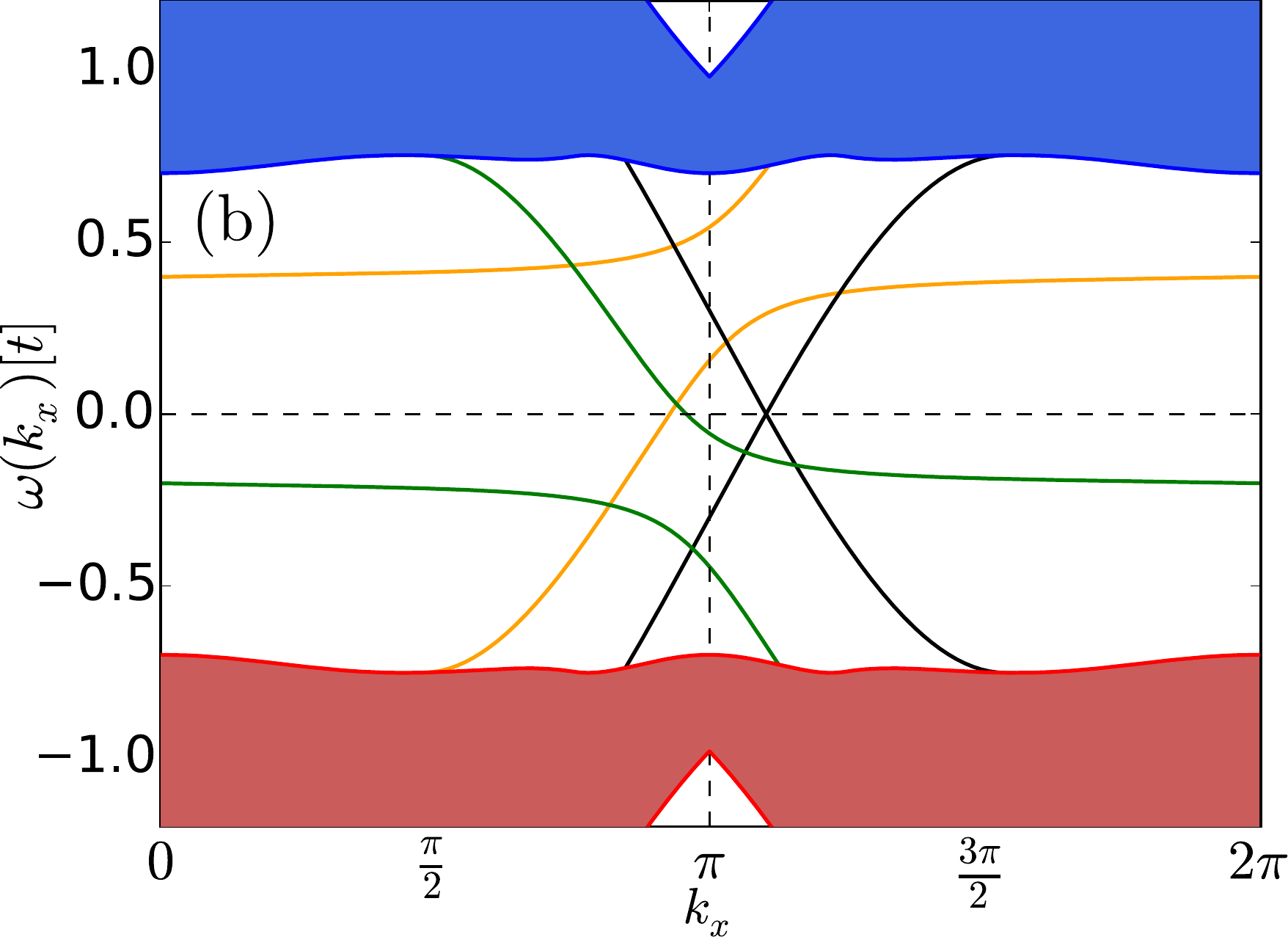}
	\caption{(Color online) (a) Dispersion of the edge states at 
	\smash{$t_2 = 0.2 t$},$t_\mathrm{r} = 0$, $\lambda_\mathrm{t} = 0.2 t$, and 
	$\delta_\mathrm{t} = 0.1 t$. The 	filled areas indicate the continua of the 
	bulk states. The edge states located at 
	the top edge are shown in color (gray).
	The spin $\uparrow$ mode propagating to the right is marked in orange 
	(light gray) while the spin $\downarrow$ mode counterpropagating to the left 
	is marked in green (dark gray). 
	The dispersion of the edge states at the lower boundary are displayed in black.
		(b) Dispersion of the edge states at \smash{$t_2 = 0.2 t$}, 
		$t_\mathrm{r} = 0$, $h_z = 0.3 t$ in Eq.\ \eqref{eq:FMX}, 
		$\lambda_\mathrm{t} = 0.2 t$, and $\delta_\mathrm{t} = 0.1 t$.}
	\label{fig:kanemele_sz}
\end{figure*}

The inclusion of a finite Rashba coupling $t_\mathrm{r}\neq 0$ violates the 
$S^z$ conservation and the two Haldane models hybridize. 
The Rashba coupling alone without the imaginary NNN hopping
does not lead to a topologically nontrivial phase \cite{berne13} 
which means that the imaginary NNN hopping is indispensable for the 
anomalous QSHE in the Kane-Mele model. But
the Rashba coupling reduces or enhances the bulk gap and influences
the edge states in this way.  An exemplary 
dispersion of the helical edge states in the Kane-Mele model is 
depicted in Fig.\ \ref{fig:kanemele_ras} where the Rashba 
coupling $t_\mathrm{r}$ has been chosen fairly large in order to 
show its influence on the bulk and on the edge states. 
For not too large values of the Rashba coupling
the qualitative features of the bulk and of the 
helical edge states remain unaltered.


The counterpropagating edge modes forming a Kramer's pair still cross each other 
as long as the TRS is preserved and the bulk gap does not close. If the gap is reduced by tuning $t_\mathrm{r}$ the bulk states repel the edge modes. 
The effect can be seen in  Fig.\ \ref{fig:kanemele_ras} where the energies of the
Kramer's pair at the bottom edge are shifted downwards. 
As a result, $v_{\mathrm{F}}$ can increase or decrease 
upon switching on the Rashba coupling as shown in Fig.\ \ref{fig:rashba}. 
Since the particle-hole symmetry is broken by the Rashba coupling the 
inclusion of the local potentials at the decorating sites is 
no longer symmetric so that the effect of a negative potential differs from the one of a positive potential. Even the sign of the effect can change.

In a system preserving TRS the addition of a finite amount of unpolarized charge
at one edge does not lead to a net charge current because the two
counterpropagating modes compensate in charge due to their equal 
Fermi velocities. In order to create a net charge current  the TRS 
must be broken. One possible way is to include a spin-dependent decoration. 
This can be accomplished for example by 
proximity-induced ferromagnetic exchange at the interface with a 
magnetic insulator \cite{jiang16}. To demonstrate this basic idea we replace 
$\mathcal{H}_{\mathrm{decor}}$ by
\begin{eqnarray}
	\mathcal{H}_{\mathrm{decor}} &=& \sum_{i, \gamma, \alpha} \left[ \lambda_{\gamma}  
	\left(c^{\dagger}_{d(i)\alpha} c_{i\alpha}^{\phantom{\dagger}} + c^{\dagger}_{i\alpha} c_{d(i)\alpha}^{\phantom{\dagger}} \right) 
	\right. 
	\nonumber\\ 
	\phantom{\mathcal{H}_{\mathrm{decor}}} && \phantom{\sum_{i \gamma \alpha}} \, \, \left. + 
	\delta_{\gamma} c^{\dagger}_{d(i)\alpha} \sigma^{z \phantom{\dagger}}_{\alpha\alpha} c_{d(i)\alpha}^{\phantom{\dagger}} \right].
	\label{eq:kane_deco2}
\end{eqnarray}

The change relative to Eq.\ \eqref{eq:kane_deco} is that 
the local potential depends on the Pauli matrix $\sigma_z$. To 
illustrate the difference to the previous decoration we depict the resulting dispersion in Fig.\ \ref{fig:kanemele_sz}(a) 
keeping all other parameters as before. 
Due to the spin-dependent decoration of the top edge the 
corresponding Kramer's doublets do no longer exist. Furthermore, 
the counterpropagating edge modes do not cancel each other. 
Hence, a net charge and spin current is possible.

Another possible way to break the TRS is to split 
the two spin states by adding a ferromagnetic exchange field 
\begin{equation}
\label{eq:FMX}
\mathcal{H}_\mathrm{FMX} = h_z\sum_{i,\alpha} (c_{i\alpha}^{\dagger} \sigma^z_{\alpha\alpha} 
c_{i\alpha}^{\phantom{\dagger}} +  c_{d(i)\alpha}^{\dagger} \sigma^z_{\alpha\alpha} c_{d(i)\alpha}^{\phantom{\dagger}} )
\end{equation}
 to the decorated model in Eq.\ \eqref{eq:kane_mele}. 
In contrast to the previous example, the exchange field 
is present at all sites. This may be realized by magnetic doping 
\cite{jungw06,qiao10,chen11b}.
For vanishing Rashba coupling $t_\mathrm{r} = 0$, the influence of the exchange field can be easily understood  by regarding the Kane-Mele model as 
two decoupled decorated Haldane models 
of which the chemical potentials are shifted in the opposite directions. 
Kramer's doublets do no longer exist, see Fig.\ \ref{fig:kanemele_sz}(b). 

\section{Robustness of the edge states against potential disorder}
\label{sec:robustness}


So far, we analyzed how an important transport property of the edge states, the Fermi velocity, can be controlled by tuning parameters. 
But there are also uncontrollable properties of 
a solid state system. For instance, imperfections of all kinds such as 
impurities, defects or vacancies in the lattice structure can never be fully excluded. We cannot consider them exhaustively here. 
But we aim at a first study of the robustness of
the edge states with respect to disorder.  To this end, we consider 
disorder in the local potentials. 

\begin{figure}
	\centering
		\includegraphics[width=1.0\columnwidth]{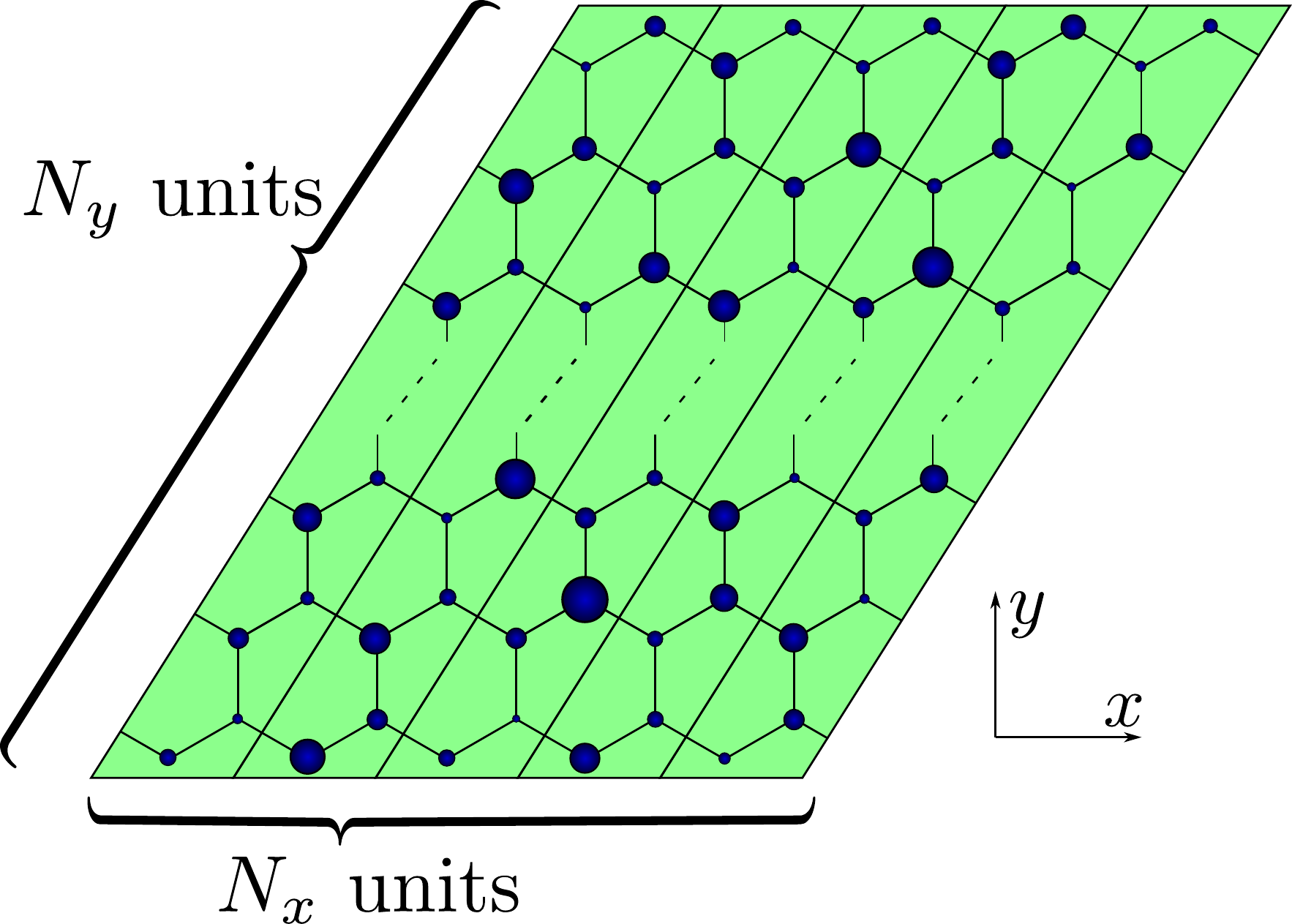}
	\caption{(Color online) Sketch of a finite strip of honeycomb lattice 
	consisting of $N_x$ columns of finite width of $N_y$ units. Each column is enclosed by thin black lines. The different sizes of the dots
	illustrate the random local potentials of a disorder configuration.}
	\label{fig:disorder}
\end{figure}

The edge states emerge as a result of the discontinuity 
of topological invariants at the edges of a system. Since a topological invariant
is a  global property of the bulk system it is expected that the edge
states are protected  as long as the disorder does not 
change the global properties of the bulk system. We want to study this explicitly.
To this end, we investigate the Haldane model \eqref{eq:haldane} on a finite strip of the honeycomb lattice as shown in Fig.\ \ref{fig:disorder}. 
We consider a strip of $N_x$ columns of a finite width of $N_y$ units so that
there are $2N_x N_y$ sites. We add a random local
 potential at each  site to the Haldane model \eqref{eq:haldane}
 to simulate the disorder. The 
random energies are taken from a continuous uniform distribution in the interval 
$\left[- \sqrt{3} \sigma, \sqrt{3} \sigma \right]$ where the 
standard deviation is given by $\sigma$.  This is the control parameter for 
the strength of the local disorder. 
We also investigated random local potentials which are 
Gaussian distributed, but the results do not differ fundamentally.

The translation symmetry in the $x$ direction is no longer preserved due to disorder.
In order to establish a link to the system without disorder we continue to
consider periodic boundary conditions. By diagonalizing the $(2 N_y N_x) \times 
(2 N_y N_x)$ matrix encoding hopping and local energies we obtain the eigenenergies.
The corresponding eigenvectors cannot be classified directly according to their
momenta $k_x$. The eigenstates are given in spatial representation by
\begin{equation}
\label{eq:realspace}
\ket{\psi} = \sum_{x, y} c(x, y) \ket{x, y} ,
\end{equation}
where $x$ and $y$ correspond to the discrete coordinates of the lattice sites.
In order to map the eigenstates of the disordered system to the eigenstates
of the clean system we express the edge states, right- and left-moving ones, of
the clean system in real space. Denoting
the wave function of an edge state by $\ket{\psi_{\mathrm{cl}}}(k_x)$
it reads
\begin{subequations}
\begin{eqnarray}
\ket{\psi_{\mathrm{cl}}}(k_x) &=& \sum_{y} d(k_x, y) \ket{k_x, y} 
\nonumber \\ 
&=& \sum_{x, y} d(k_x, y) \frac{\mathrm{e}^{- \mathrm{i} k_x x}}{\sqrt{N}} 
\ket{x, y} \, .
\end{eqnarray}
In comparison to the representation \eqref{eq:realspace} we deduce
\begin{equation}
c_{k_x}(x,y) = d(k_x, y) \frac{\mathrm{e}^{- \mathrm{i} k_x x}}{\sqrt{N}}.
\end{equation}
\end{subequations}
The possible momenta are given by $k_x = 2 \pi n_x/N_x$ with 
$n_x = \left\lbrace 0, 1, \ldots,  N_x - 1 \right\rbrace$.

In order to assign a momentum $k_x$ to an energy of an edge state of
the disordered system we search for the largest overlap with a clean
edge mode, i.e., we maximize 
$\left| \braket{\psi}{\psi_{\mathrm{cl}}} \right|^2 (k_x)$
by varying $k_x$. The momentum $k_x$ which maximizes this overlap is the one assigned to the eigenstate of the disordered system.
The overlap can be interpreted as transition probability and is calculated by
\begin{equation}
\label{eq:trans-probabl}
\left| \braket{\psi}{\psi_{\mathrm{cl}}} \right|^2 (k_x) 
= | \sum_{x,y} c^*(x,y)c_{k_x}(x,y) |^2 .
\end{equation}

\begin{figure}
	\centering
		\includegraphics[width=1.0\columnwidth]{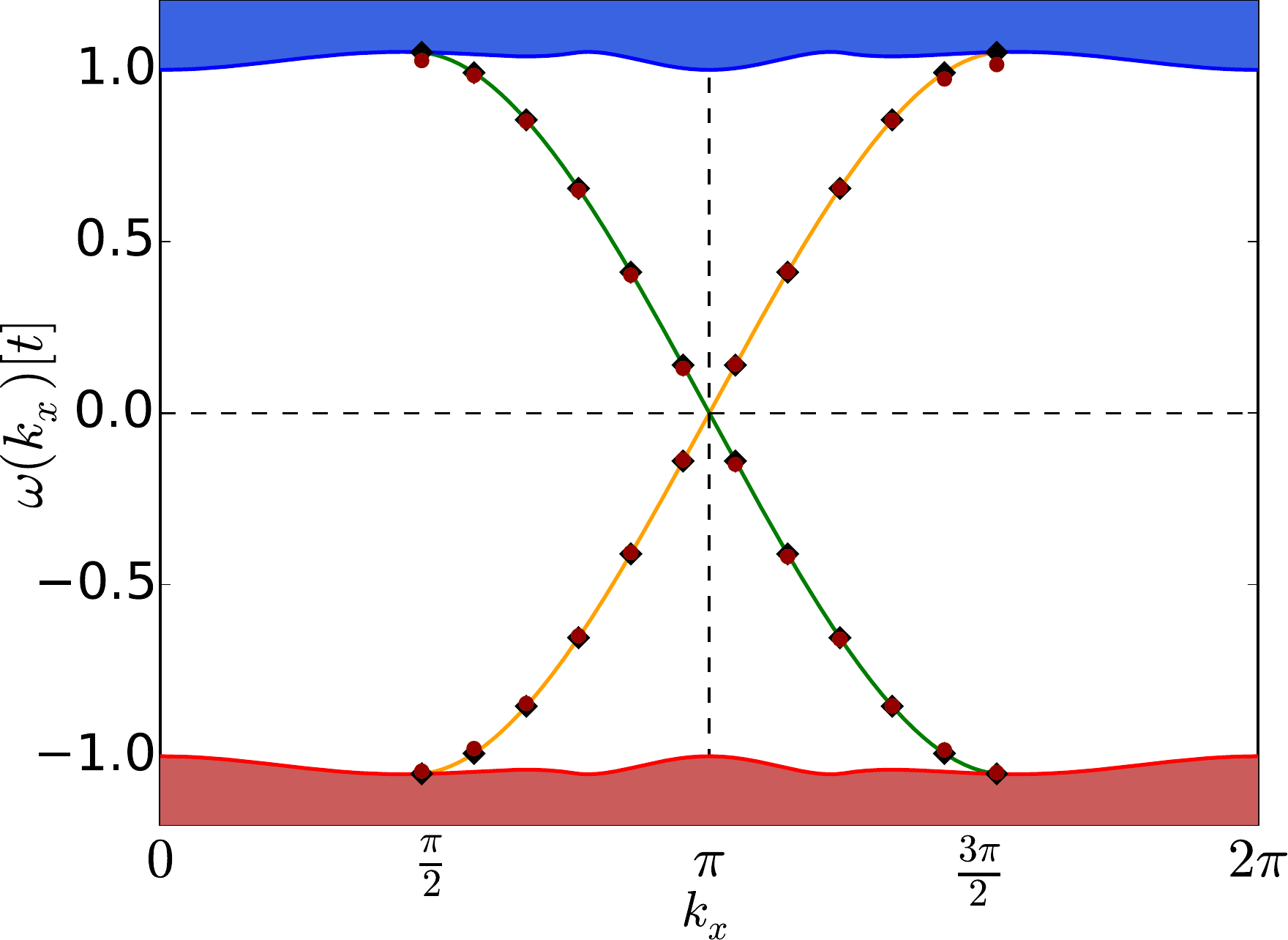}
	\caption{(Color online) Dispersion of the edge states with $t_2 = 0.2 t$ and 
	$\phi = \nicefrac{\pi}{2}$ of the Haldane model. The filled 
	areas indicate the continua of the
	bulk states. The right-moving edge state marked in orange (light gray) 
	is located at
	the top edge while the left-moving edge state marked in green (dark gray) is
	located at the bottom edge. The symbols indicate the energies of eigenstates 
	of which	the momenta are determined from maximizing the transition 
	probability in 
	\eqref{eq:trans-probabl}. The black diamonds are calculated 
	for a clean system with $N_y = 50$ and $N_x = 21$ while the red (gray) circles 
	result from a disordered system with $\sigma = 0.1 t$.}
	\label{fig:edge}
\end{figure}

Following this procedure, we reconstruct the dispersion of the edge state in the Brillouin zone as shown in \smash{Fig.\ \ref{fig:edge}} for $\sigma = 0.1 t$. Typically, we consider a system of 
$N_y = 50$ and $N_x = 21$ leading to  
$2 N_y N_x = 2100$ eigenenergies from which we
 select the energies corresponding to the edge state by maximizing
the transition probability \eqref{eq:trans-probabl}. 
The dispersions of the edge modes of the clean system computed from the infinite strip ($N_x=\infty$)  are shown as solid lines for the sake of comparison. 
To test the maximization of the transition probability we 
assign momenta to eigenstates computed for a finite clean system. The results
are depicted by black diamonds in \smash{Fig.\ \ref{fig:edge}} and match
the continuous lines perfectly.
The red (gray) circles display the eigenenergies at the assigned momenta
in a disordered system with $\sigma = 0.1 t$. They are still
located close to the solid lines, but do not match them perfectly due
to the disorder. 

We conclude that the qualitative features of the edge
states are indeed robust against disorder. The gaplessness of the edge
modes is preserved as was to be expected from the topological protection.
But also the quantitative aspects are not drastically altered by disorder,
at least as long as the disorder strength is not too large.

\begin{figure}
	\centering
		\includegraphics[width=1.0\columnwidth]{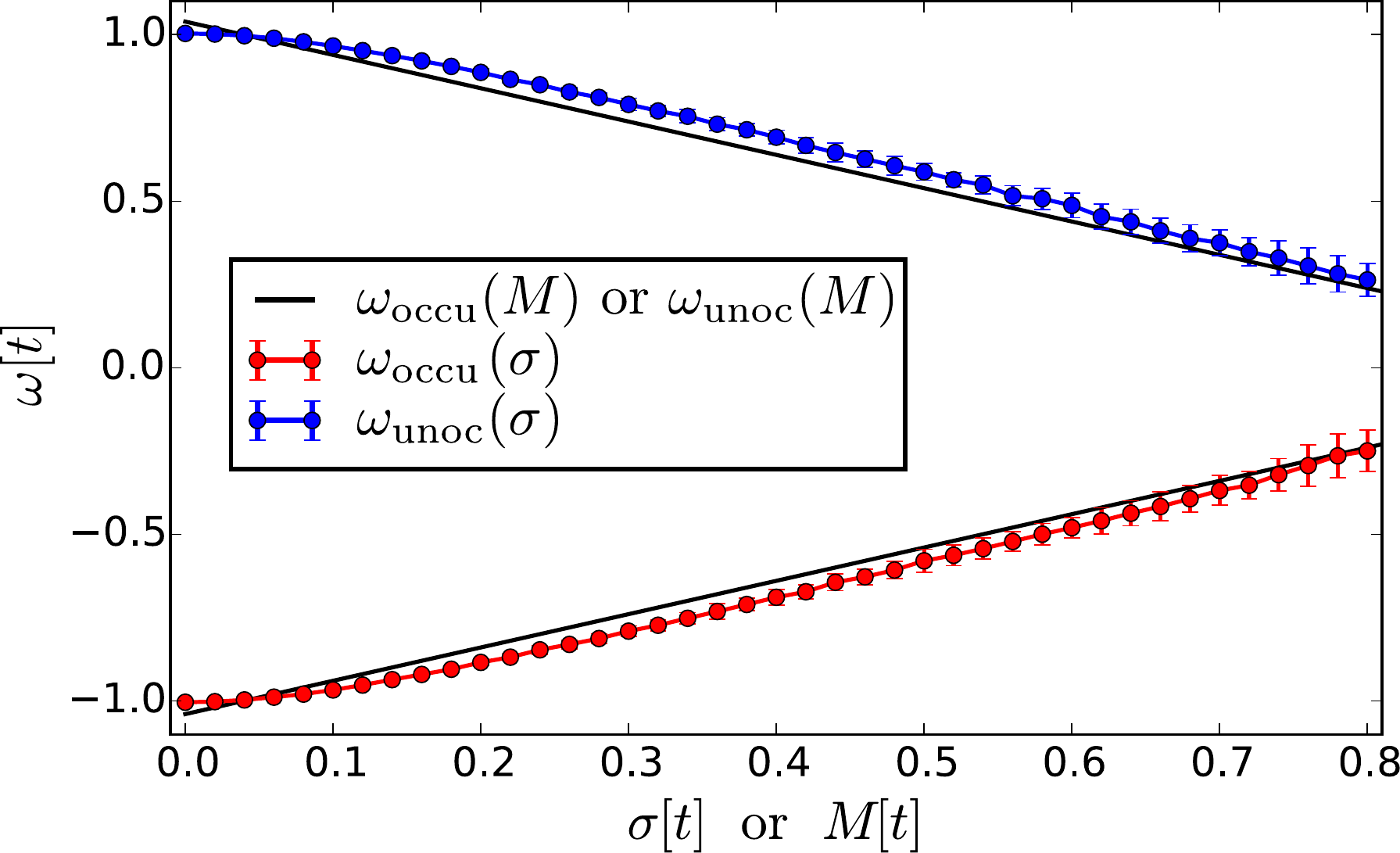}
	\caption{(Color online) The lower band edge of the conduction band 
	(blue, dark gray) 
	and the maximum energy of the valence band (red, light gray) vs.\ the
	disorder strength $\sigma$ in a system with $N_y = 50$ 
	and $N_x = 21$. The energies are averaged over 60 randomly chosen configurations. 
	The error bars represent a standard deviation. 
	The black lines show the band edges in a clean system 
	as a function of a local inversion-symmetry breaking term $\propto M$, 
	see main text.}
	\label{fig:energy_gap}
\end{figure}

An important point to study is the influence of the disorder on the bulk gap. 
If the bulk gap becomes small or even vanishes the topological properties disappear.
Increasing disorder reduces the bulk gap.  An estimate for this reduction
can be derived by assuming that the disorder strength $\sigma$ 
behaves similar to an on-site inversion-symmetry breaking term 
$\varepsilon_i^{\phantom{\dagger}} M c_i^{\dagger} c^{\phantom{\dagger}}_i$.
Here $\varepsilon_i$ takes the values  $\pm 1$ depending on whether 
site $i$ belongs to one sublattice or to the other. 
The energy gap $\Delta$ of the bulk system decreases upon increasing $M$.
Similarly, $\Delta$ decreases upon increasing $\sigma$ as we illustrate
in \smash{Fig.\ \ref{fig:energy_gap}} where the lower band edge 
$\omega_\text{unoc}=\Delta/2$ of the unoccupied
states and the upper band edge 
$\omega_\text{occu}=-\Delta/2$ of the occupied states are shown.
 The black solid lines depict the bulk gap as a function of $M$ according to 
\smash{$\Delta = 2 | M \pm 3 \sqrt{3} t_2 \sin{\phi}|$} 
\cite{halda88b,fruch13,chen11b}. The symbols show the bulk gap
in the disordered sample determined in the following way.
For the lower band edge we compute the minimum energy of the eigenstates which
\emph{cannot} be assigned to an edge mode of the clean system. Similarly, the upper
band edge is determined from the maximum energy of the eigenstates which
\emph{cannot} be assigned to an edge mode of the clean system.
Of course, this way of determining the bulk gap in the disordered system
is a heuristic one and not mathematically rigorous.
But the comparison to $\omega_{\mathrm{occu}}(M)$ and $\omega_{\mathrm{unoc}}(M)$ shows good agreement so that
we conclude that the estimate works very well.

The energy gap disappears at $M = 3 \sqrt{3} t_2 \sin{\phi}$ 
\cite{halda88b,fruch13,chen11b}. 
Thus, the estimate predicts that the topological properties will
definitely cease to exist for a disorder strength
\begin{equation}
\sigma \approx  3 \sqrt{3} t_2 \sin \phi \, . 
\end{equation}
We stress that the decreasing bulk gap reduces the energy interval in which
the edge mode can be identified. Concomitantly, the interval in momentum $k_x$
in which the edge mode can be identified is reduced as well. 

\begin{figure}
	\centering
		\includegraphics[width=1.0\columnwidth]{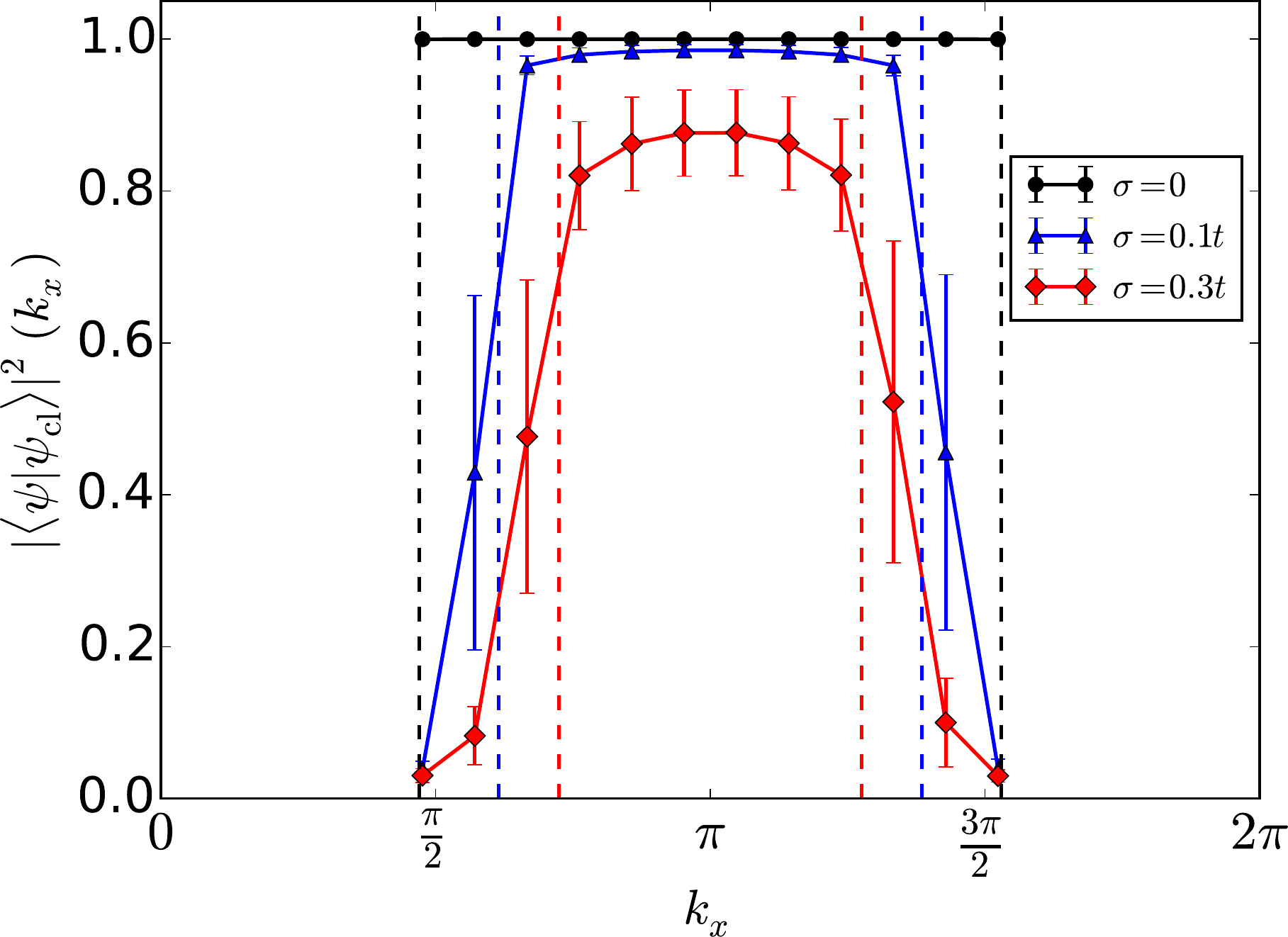}
	\caption{Transition probability 
	$\left| \braket{\psi}{\psi_{\mathrm{cl}}} \right|^2 (k_x)$
	of the right-moving edge state 
	averaged over $50$ random configurations in a system with 
	$N_y = 50$ and $N_x = 21$
	as a function of the momentum $k_x$. The error bars indicate the 
	standard deviation. The dashed 
	lines indicate at which momenta the energy of the edge mode enters the
	bulk continua.}
	\label{fig:kx}
\end{figure}

Next, we study how well the edge mode can be identified close to 
the bulk continua. \smash{Figure \ref{fig:kx}} displays the transition probability
 $\left| \braket{\psi}{\psi_{\mathrm{cl}}} \right|^2 (k_x)$ 
of the edge states in the Brillouin zone. The vertical dashed lines 
indicate the thresholds where the edge modes enter the bulk continua, i.e., 
where the energies of the edge modes exceed the estimated bulk gap.
It is obvious that around $k_x=\pi$ the transition probability
between the edge mode in the disordered system and in the clean system
is large. Thus, in particular for low disorder, the identification 
of the edge mode works reliably. For increasing disorder, the overlap decreases
gradually. Approaching the band edges at fixed disorder strength, 
i.e., approaching the dashed line, the overlap
decreases rapidly and a clear identification of the edge modes becomes
more and more difficult untill it becomes impossible. This data shows
the breakdown of the edge modes under the influence of disorder.
Clearly, there are limits to the topological protection, even though
the feature of a vanishing energy of the edge modes persists as required.

\begin{figure}
	\centering
		\includegraphics[width=1.0\columnwidth]{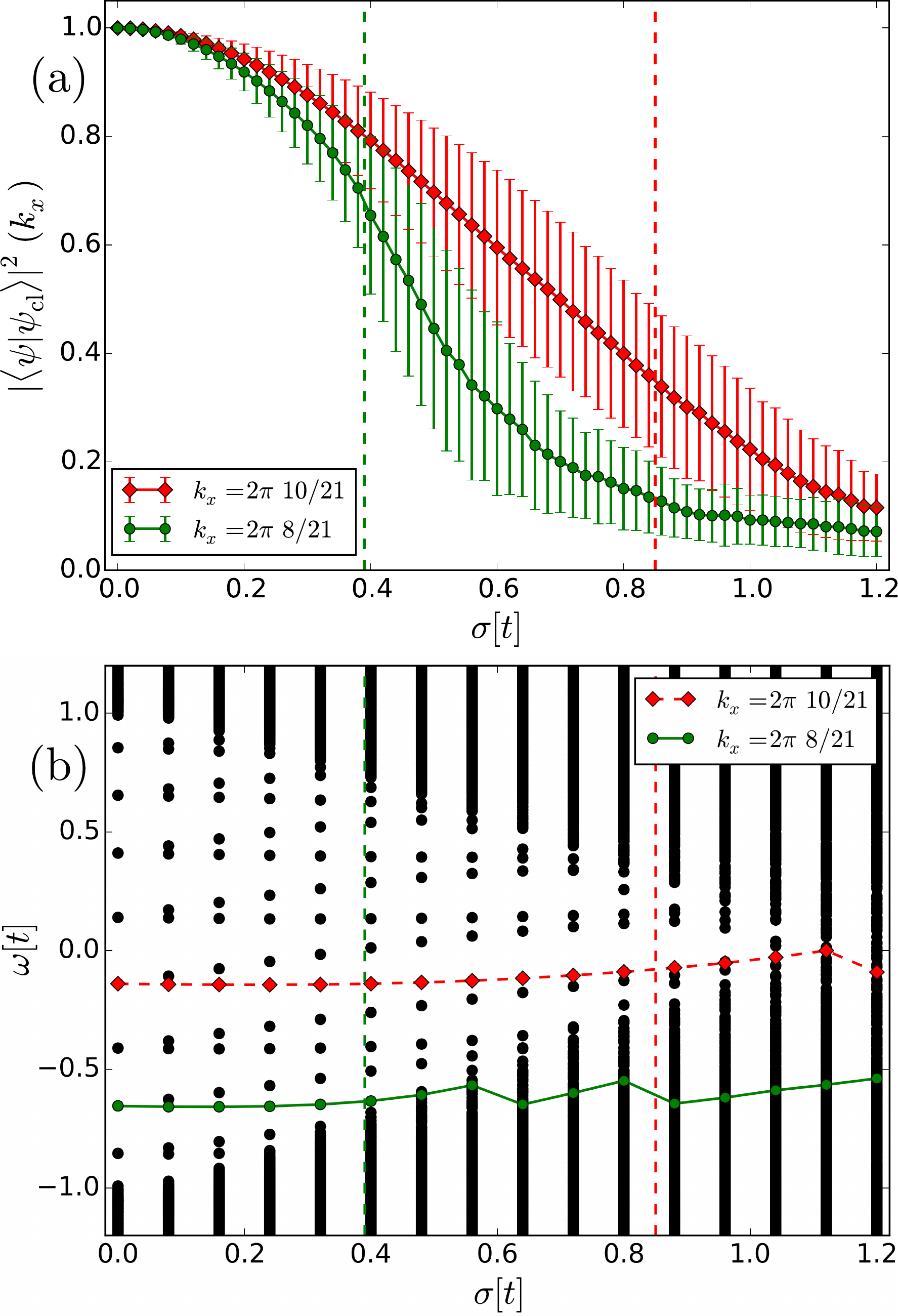}
	\caption{(a) Transition probability
	$\left| \braket{\psi}{\psi_{\mathrm{cl}}} \right|^2 (k_x)$ at two values of
	$k_x$ for the right-moving edge state as a function of the disorder strength 
	$\sigma$ in a system with 	$N_y = 50$ and $N_x = 21$. The probability is  
	averaged over $50$ configurations. The error bars represent the standard deviation. The dashed lines depict where the edge modes 
	enter the continuum of the bulk states. 
	(b) Complete energy spectrum for an exemplary configuration. 
	The energies highlighted in color correspond to the two values of 
	$k_x$  displayed in (a).}
	\label{fig:sigma}
\end{figure}


The quantitative behavior of 
$\left| \braket{\psi}{\psi_{\mathrm{cl}}} \right|^2 (k_x)$ as a 
function of $\sigma$ is studied in \smash{Fig.\ \ref{fig:sigma}(a)}. The transition probability decreases upon increasing $\sigma$.  
Beyond a certain value \smash{of $\sigma$} the transition
probability  saturates at a small residual value. If the energy of the edge mode
in the clean system is far away from the band edges of the continua
 (red curve, symbol 1)
the transition probability decreases more slowly than 
if its energy is close
to one of the continua (green curve, symbol 2). 

In  \smash{Fig.\ \ref{fig:sigma}(b)} we depict the dependence of the 
complete energy spectrum on the disorder strength. 
The modes assigned to the two momenta shown in
\smash{Fig.\ \ref{fig:sigma}(a)} are highlighted by the two lines. There are
 regions where the eigenenergies are dense corresponding to the continua.
The energies between the two dense regions at low and at high energies
belong to the edge modes. The energies assigned to the
two momenta evolve upon increasing $\sigma$. At some
value of $\sigma$, which is specific for the momentum $k_x$ of the mode, they
enter the bulk continuum. The corresponding values \smash{of $\sigma$} are 
indicated approximately by vertical dashed lines 
in both panels of \smash{Fig.\ \ref{fig:sigma}}.
Beyond these disorder strengths it can no longer be decided whether the
modes are true edge modes or whether they belong to the continuum states.

Yet even beyond the dashed lines the transition probability is
large enough to assign energies to the \smash{momenta $k_x$}. But 
it happens that the assigned  energies jump as can be seen 
for $k_x = 2 \pi \, 8/21$ where kinks occur beyond the dashed line.
This indicates that the assignment energy $\leftrightarrow$ momentum
based on the transition probability is no longer reliable.

\begin{figure}
	\centering
		\includegraphics[width=1.0\columnwidth]{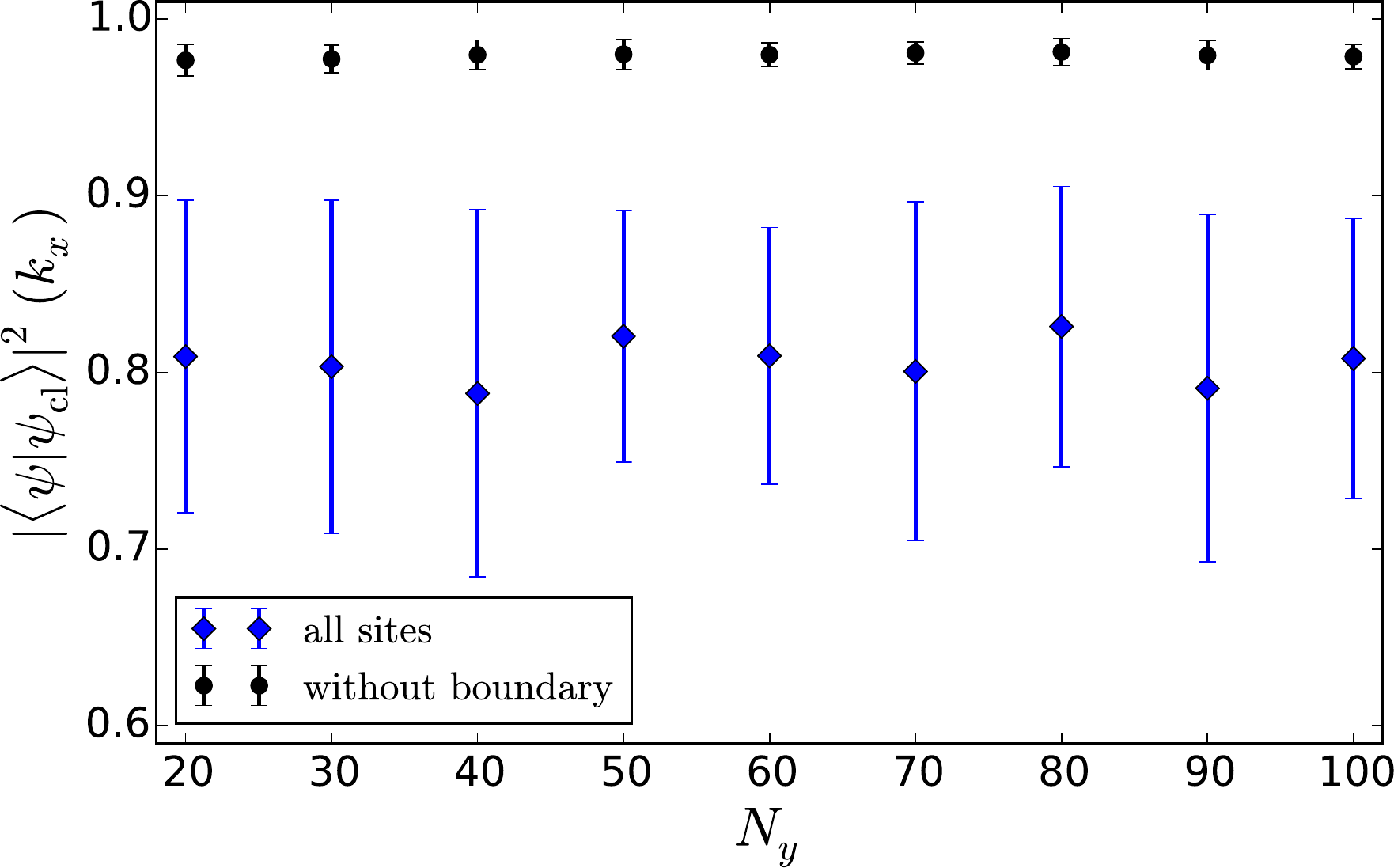}
	\caption{(Color online) Transition probability 
	$\left| \braket{\psi}{\psi_{\mathrm{cl}}} \right|^2$
	at $k_x = 2 \pi \, 8/21$ as a function of the width $N_y$ of the strip
	of the Haldane model with $t_2 = 0.2 t$, 
	$\phi = \pi/2$, and a length of $N_x = 21$. 
	The probability is averaged over $50$ configurations for
		$\sigma= 0.1 t$. The error bars represent a standard deviation. The effect
		of local disorder on all sites is shown by black circles. 
		The blue (gray) circles depict the effect if disorder is only present 
		in the bulk, but not at the edges.
		As to be expected, the edge modes are much less influenced in this case.}
	\label{fig:ny}
\end{figure}

Next, we address the dependence of the modes on the width $N_y$ and the 
length $N_x$ of the system. Larger $N_y$ increases the width of the 
strip. Since the edge modes are localized at the boundaries 
increasing the width separates them more and more and makes them 
independent from each other. We focused on wide enough strips anyway
so that the edge modes are essentially independent of $N_y$. 
This is supported clearly by \smash{Fig.\ \ref{fig:ny}}. The width $N_y$ of 
the strip plays no important role once it is large enough.
Furthermore, \smash{Fig.\ \ref{fig:ny}} shows that 
\smash{$\left| \braket{\psi}{\psi_{\mathrm{cl}}} \right|^2 $} crucially depends on the characteristics of the edges. If the edges are unaffected by the local 
disorder the transition probability takes significantly larger values 
than in the case where all
sites are subject to random potentials.

\begin{figure}
	\centering
		\includegraphics[width=1.0\columnwidth]{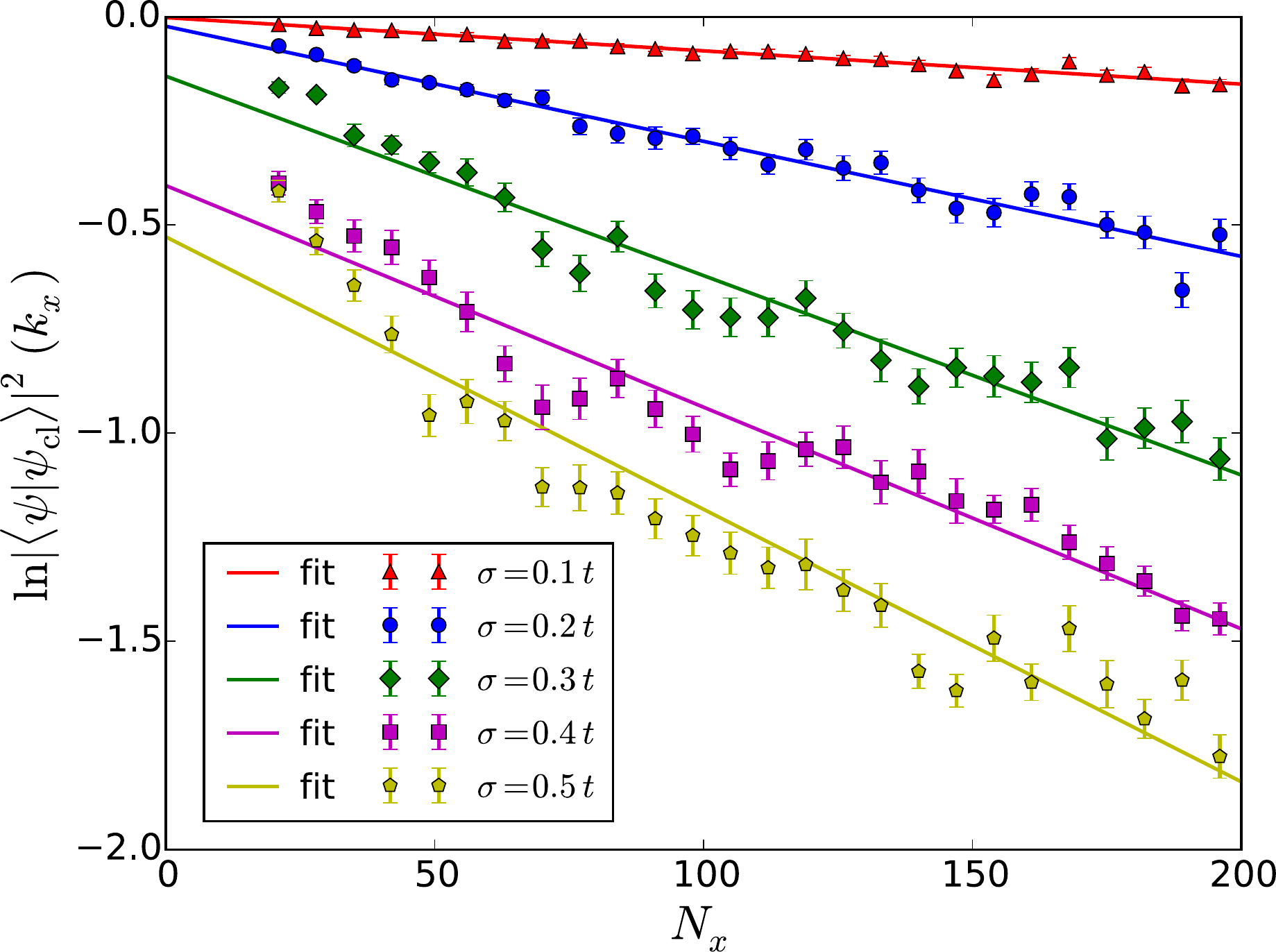}
	\caption{(Color online) The logarithm of the transition probability at 
	the momentum $k_x = 2 \pi \, 10 / 21$ vs.\ the length $N_x$ of the system at 
	$t_2 = 0.2 t$,  $\phi = \pi/2$, and the width $N_y = 20$. 
	The transition probability is averaged over $50$ random configurations 
	for various disorder strengths $\sigma$. 
	The error bars represent the standard deviation of the average.}
	\label{fig:nx}
\end{figure}

Increasing the length $N_x$ of the system has a pronounced effect on the
transition probability as shown in \smash{Fig.\ \ref{fig:nx}}. Note the logarithmic
scale of the $y$-axis. Though the numerical data for the transition probability
 $\left| \braket{\psi}{\psi_{\mathrm{cl}}} \right|^2 (k_x)$ is a bit noisy, it
agrees well with an exponential dependence 
\begin{equation}
\left| \braket{\psi}{\psi_{\mathrm{cl}}} \right|^2 
(k_x) \propto \exp(-\gamma(\sigma) N_x)
\end{equation}
where the rate $\gamma$ depends on the disorder strength. Naturally, the 
overlap decreases more rapidly if the disorder strength is larger.

The observed dependence on $N_x$ can be understood as follows. Let us view
a system of given length $N_x$ to be formed by concatenating a number $r$ of 
short subsystems of length $n_x$ with $N_x=r\cdot n_x$. 
If the subsystems are still long enough, the physics inside of each of them
is only negligibly influenced by the boundaries between them. Then, the
transition probability of the total system is given by the product of
all the transition probabilities of the subsystems
\begin{equation}
\left| \braket{\psi}{\psi_{\mathrm{cl}}}(N_x) \right|^2 
=\prod_{j=1}^r \left| \braket{\psi}{\psi_{\mathrm{cl}}}(n_x,j) \right|^2 .
\end{equation}
On average, the transition probabilities of all the subsystems are the same
so we denote them by $p_\mathrm{sub}<1$. Thus we have
\begin{subequations}
\begin{eqnarray}
\left| \braket{\psi}{\psi_{\mathrm{cl}}}(N_x) \right|^2 
&=& p_{sub}^r \\
&=& \exp(-\tilde\gamma r)\\
&=&  \exp(-\gamma N_x) 
\label{eq:exp}
\end{eqnarray}
\end{subequations}
where we set $p_\mathrm{sub}=\exp(-\tilde\gamma)$ and $\gamma=\tilde\gamma/n_x$.

Inspecting \smash{Fig.\ \ref{fig:nx}} reveals that the exponential decay 
does not apply for short systems, but only beyond a 
certain minimum length. Thus, the above 
argument is only approximately true because the assumption of negligible influence
of the boundaries is not perfectly justified for short systems. 
Thus a linear fit $a - \gamma \, N_x$ 
of the logarithm of $\left| \braket{\psi}{\psi_{\mathrm{cl}}}(N_x) \right|^2$
as shown in \smash{Fig.\ \ref{fig:nx}} works well, but the offset $a$ is not zero
in contrast to what our simple argument suggests in Eq.\ \eqref{eq:exp}.
The fitted values are given in Table \ref{tab:nx}. 

\begin{table}
\centering
\begin{tabular}{c @\quad c @\quad c}\toprule
$\sigma$ & $a \pm \Delta a$ & $\gamma \pm \Delta \gamma$\\ \cline{1-3}
$0.1$ & $-0.00174 \pm 0.00458$ & $0.00080 \pm 0.00004$ \\
$0.2$ & $-0.02261 \pm 0.01491$ & $0.00276 \pm 0.00012$ \\
$0.3$ & $-0.14242 \pm 0.02713$ & $0.00479 \pm 0.00023$ \\
$0.4$ & $-0.40515 \pm 0.03240$ & $0.00532 \pm 0.00027$ \\
$0.5$ & $-0.52846 \pm 0.04800$ & $0.00654 \pm 0.00040$ \\
\botrule
\end{tabular}
\caption{Fitted values of the linear fits 
$\ln \left| \braket{\psi}{\psi_{\mathrm{cl}}}(N_x) \right|^2\approx 
a - \gamma \, N_x$ in \smash{Fig.\ \ref{fig:nx}}.}
\label{tab:nx}
\end{table}

Finally, we study the effect of disorder on the edge mode at a decorated edge. 
It has been advocated that the decoration and a tunable gate voltage shifting
the potential at the edges render the realization of tunable, direction-dependent
delay lines possible \cite{uhrig16}. 
If we recall the extension to the Kane-Mele model
a dependence on the spin is also possible. This makes the fundamental
idea interesting  for spintronics as well. But for all applications the robustness
towards imperfections is decisive. This motivates the investigation of disorder.

\begin{figure*}
	\centering
	\includegraphics[width=1.0\columnwidth]{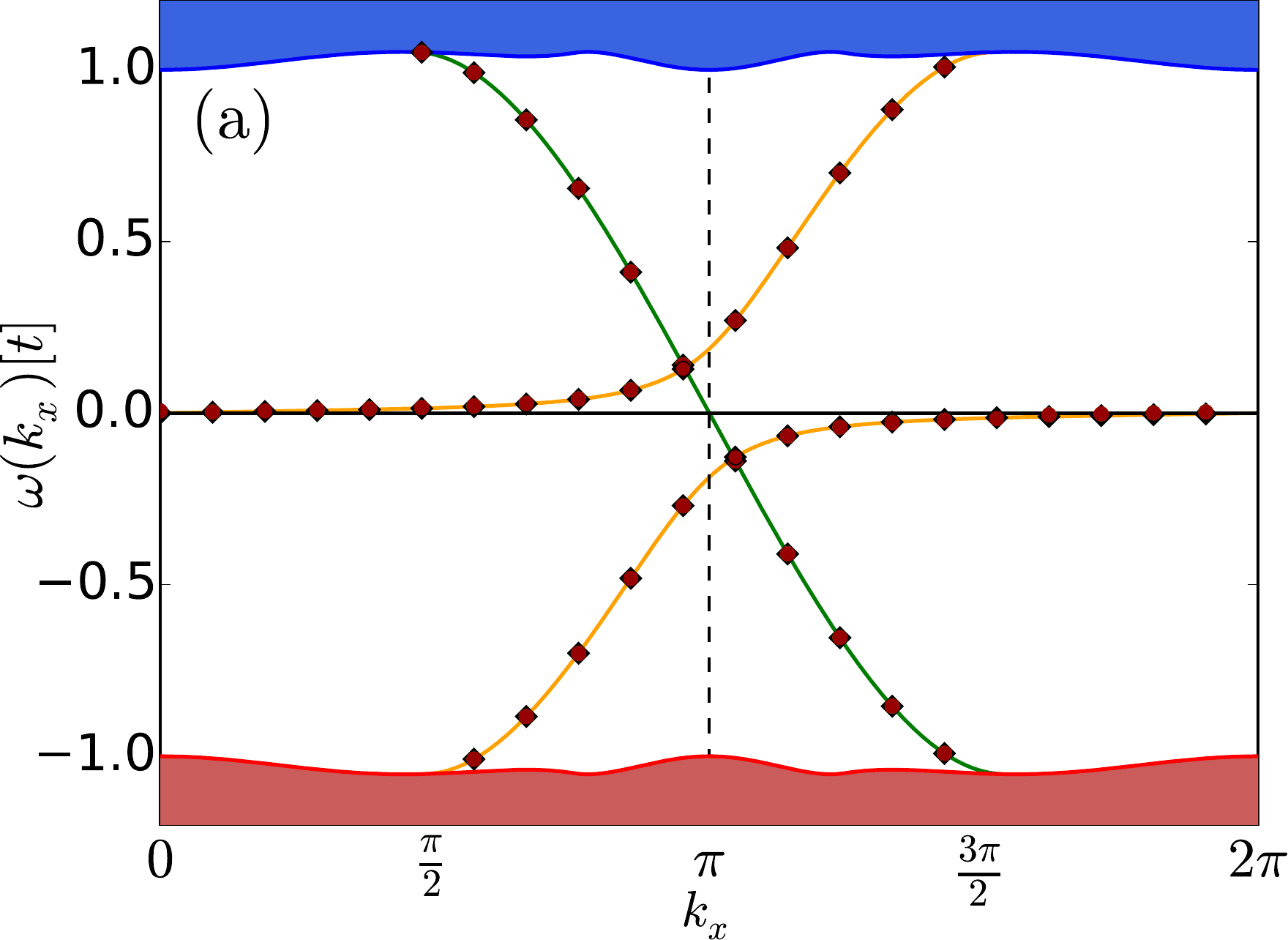}
	\includegraphics[width=1.0\columnwidth]{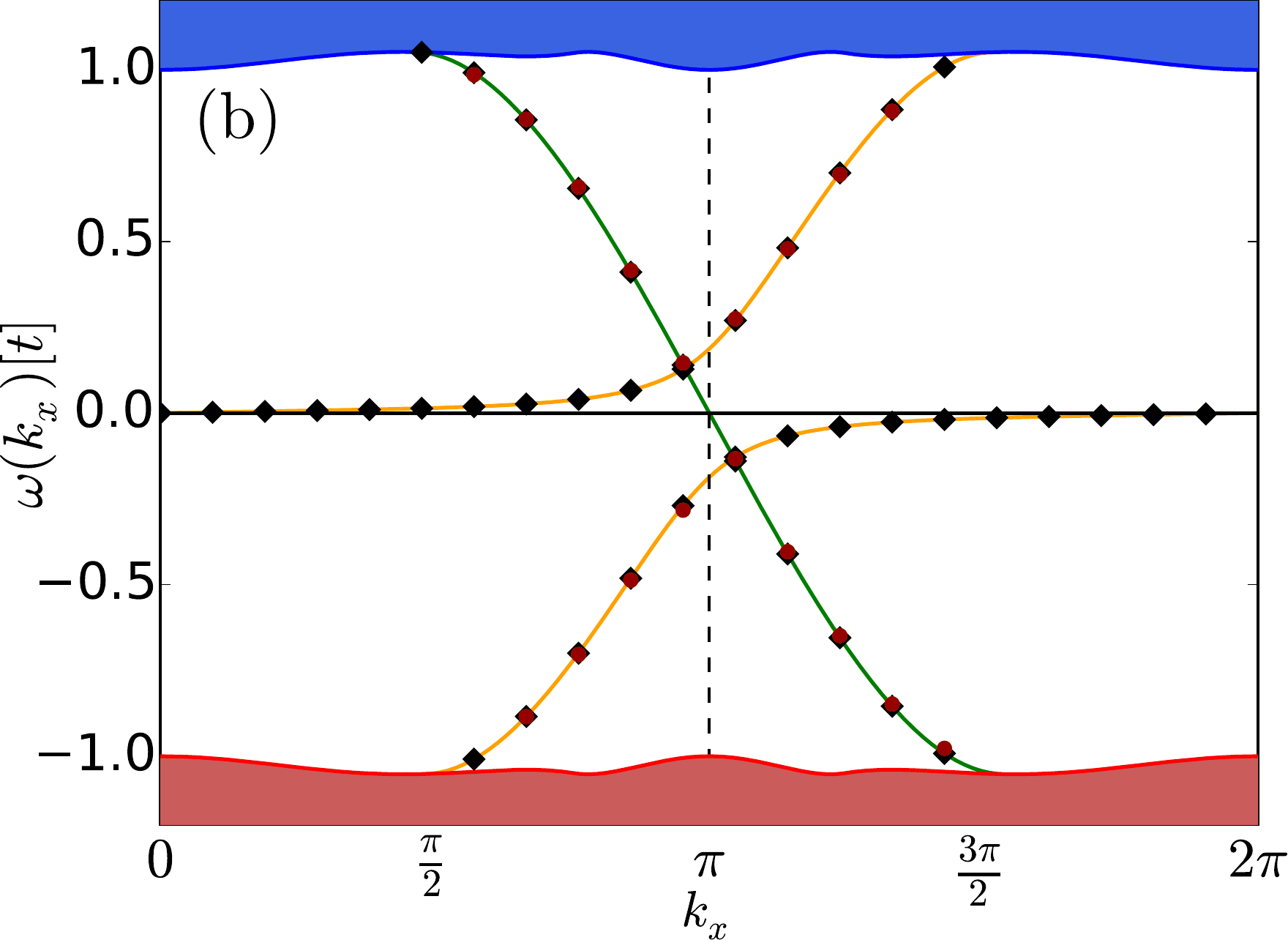}	
	\caption{(Color online) Dispersion of the edge states in 
	a decorated system with $t_2 = 0.2 t$, $\lambda_\mathrm{t} = 0.2 t$,
	and $\phi = \pi/2$. The filled areas indicate the 
	continua of bulk states. The right-moving edge state marked in orange 
	(light gray) 	is located at the top edge while the left-moving edge state 
	marked in green (dark gray) is located at the bottom edge. 
	For reference, the black diamonds depict the
reconstructed dispersion in the clean system with $N_y = 50$ and $N_x = 21$ while 
the red (gray) circles depict the reconstructed dispersion for 
$\sigma = 0.001 t$ in  panel (a) and for $\sigma = 0.1 t$ in panel (b). 
Note that in panel (b) the flat
part of the dispersion could not be reconstructed 
because the transition probabilities fall below the required threshold, 
see main text.}
	\label{fig:edge_top}
\end{figure*}

The purpose of the decoration is to reduce the Fermi velocity by design, i.e.,
to introduce fairly flat regions in the dispersion. This implies that there are
many eigenstates of very similar energies. From perturbation theory one knows
that such systems are susceptible to generic perturbations such as disorder.
We investigate a system of size $N_y = 50$ and $N_x = 21$ with a decorated upper boundary. Since the decorating sites are not excluded from disorder
we also add a random local potential to the additional sites.
In the reconstruction of the dispersion of the edge modes we require a
certain minimum transition probability in order to obtain a reliable mapping
between momenta and eigenstates. From the above results for systems of the
considered size we set this threshold to $0.3$, cf.\ Figs.\ \ref{fig:kx} and  
\ref{fig:sigma}. For weak disorder the successfully reconstructed dispersion
is displayed in \smash{Fig.\ \ref{fig:edge_top}(a)}.

For stronger disorder, a complete reconstruction of the dispersion of the edge states  turns out to be impossible, see \smash{Fig.\ \ref{fig:edge_top}(b)}. 
For instance for $\sigma = 0.1 t$ ,the eigenstates with energies within the flat 
dispersion in the center of the gap cannot be mapped reliably to the corresponding momenta because their overlap falls below the threshold. As expected from our
perturbative argument, the states in the 
flatter regions of the dispersion are not particularly robust against disorder.

For a complete understanding, we also studied the case where there is no
disorder at the decorating sites. This is a realistic scenario if the
technique which creates the decorating sites is a different one from the
one growing the bulk. Clearly, this kind of disorder has much less 
detrimental effects on the edge modes, see for instance 
\smash{Fig.\ \ref{fig:ny}}. The edge modes
are rather localized at the decorating sites so that they are less
exposed to disorder. This holds in particular for the states with 
rather flat dispersion because they differ only slightly from
the completely local states on the decorating sites.
For instance the same configuration
as used in Fig.\ \ref{fig:edge_top}(b) can be reconstructed up to
much stronger disorder $\sigma = 0.5t$ if the disorder
is restricted to the bulk.

Note that the above observations do not contradict to the general idea of topological protection because there are modes at arbitrary small energies. 
But for transmitting signals
one needs a clearly defined dispersion $\omega(k_x)$ which yields the group 
velocity $\partial\omega/\partial k_x$. If this is not the case as we found
here for stronger disorder we presume that the system is not suitable for
applications based on signal transmission. This sets certain limits to
the general idea of topological protection which should not be misunderstood as
a guarantee that dispersion and group velocity are well-defined.

\section{Summary}
\label{sec:summary}

In this paper, we concentrated on the Fermi velocity of the edge states of topologically nontrivial fermionic lattice systems. The Fermi velocity is the group velocity with which signals can be transmitted through the edge states. The tunability of the edge states can be used to create delay lines based on interference, see Ref.\ \onlinecite{uhrig16}. Thus it is a measurable quantity very important for transport behavior, but which is different from DC conductivity studied previously \cite{dyke16, qiao16}.

First, in Sec.\ \ref{sec:haldane}, we presented the decorated Haldane model 
to be able to compare later to the extended Kane-Mele model and to the
decorated Haldane model with disorder.
We discussed the effects that various parameters of the decoration
have on the properties of the edge states, most notably on their dispersion. 
The Fermi velocity is direction-dependent if the different edges are 
decorated and tuned independently.

Second, in Sec.\ \ref{sec:kane_mele}, the results for the spinless
Haldane model were extended to the spinful
Kane-Mele model. In this model, the dispersions of the edge modes depend
on the combination of direction and spin. The model in its entire 
composition does not break TRS. So for each right (or left)-moving spin 
$\uparrow$ mode there is a left (or right)-moving spin $\downarrow$ mode 
with equal energy. 
The full control of the dispersions and their dependence on direction and
spin \emph{separately} can be achieved by realizing spin-dependent exchange couplings at the edges. Candidates for the realization of such terms 
in the Hamiltonian are the proximity effect of a ferromagnet in hybrid structures
or magnetic doping in the bulk of the system.
In addition, we studied the effect of Rashba coupling.

Third, in Sec.\ \ref{sec:robustness}, we addressed the effect of disorder on the 
edge states as motivated by the fundamental paradigm of topological effects that the edge states are particularly robust against any kind of perturbation. 
For clarity, we performed this study
for the spinless Haldane model. Indeed, the existence of gapless states
at the edges is guaranteed by topological protection.
But there is no guarantee for the preservation of a well-defined
dispersion of the edge modes. Thus, the transport properties are likely
to be influenced significantly by disorder.

We reconstructed the dispersion of the edge modes in disordered 
systems by comparing them with the edge modes of the clean system.
The transition probability between the
edge state in the clean system and the one in the disordered system 
served as criterion to identify the modes.
In this way, one can link the eigenstates in the disordered systems
to certain momenta and re-define a dispersion. The approach works
very well for edge states of which the energy is far away from the
continua. But if the energies approach the band edges, the
mapping becomes ambiguous so that its application is no longer reliable.
Thus, for stronger disorder only small parts of the original dispersions
can be reconstructed. Increasing the disorder even further eventually
destroys the edge modes completely. In addition, we established an approximate
formula for the reduction of the bulk gap due to disorder
in the Haldane model.

Furthermore, we clarified how the transition probability depends on the
width and the length of the system under study. The width does not have
a significant impact once the sample is wide enough so that the two edge
modes do not interact anymore. The increasing of the length leads to an 
exponential decrease of the transition probability.

Finally, we addressed the robustness of the edge states at decorated
edges which allow us to design small and tunable Fermi velocities.
Applying the reconstruction procedure we could cope with small disorder
strengths. But we found our expectation confirmed that the flat regions
of the dispersions are particularly susceptible to perturbations. 
We conclude that in order to realize and to apply the ideas of tunable group
velocities one has to resort to clean samples or, at least, to
samples were the decorating sites are not subject to disorder.
The edge modes displaying a large dispersion
and staying away from the band edges of the bulk modes are those which
are most robust to disorder.

Further studies are necessary in order to investigate the
influence of other kinds of disorder or imperfections. On the one hand, it
is conceivable that spatially correlated disorder is less harmful
to the edge modes than the completely local one we studied here.
The edge modes may flow around smoother regions of disorder 
or imperfections,  for an example on the surface of a topological insulator,
see Ref.\ \onlinecite{bauer16}.
On the other hand, imperfections such as vacancies can behave 
like a local infinite potential, i.e., having very drastic
effects on the edge modes.

Extending such investigations to other kinds of systems displaying
topological order constitutes another broad field of research.

\acknowledgments

This work was supported by the Deutsche Forschungsgemeinschaft and the
Russian Foundation of Basic Research in the International Collaborative
Research Center TRR 160.


\begin{thebibliography}{47}%
\makeatletter
\providecommand \@ifxundefined [1]{%
 \@ifx{#1\undefined}
}%
\providecommand \@ifnum [1]{%
 \ifnum #1\expandafter \@firstoftwo
 \else \expandafter \@secondoftwo
 \fi
}%
\providecommand \@ifx [1]{%
 \ifx #1\expandafter \@firstoftwo
 \else \expandafter \@secondoftwo
 \fi
}%
\providecommand \natexlab [1]{#1}%
\providecommand \enquote  [1]{``#1''}%
\providecommand \bibnamefont  [1]{#1}%
\providecommand \bibfnamefont [1]{#1}%
\providecommand \citenamefont [1]{#1}%
\providecommand \href@noop [0]{\@secondoftwo}%
\providecommand \href [0]{\begingroup \@sanitize@url \@href}%
\providecommand \@href[1]{\@@startlink{#1}\@@href}%
\providecommand \@@href[1]{\endgroup#1\@@endlink}%
\providecommand \@sanitize@url [0]{\catcode `\\12\catcode `\$12\catcode
  `\&12\catcode `\#12\catcode `\^12\catcode `\_12\catcode `\%12\relax}%
\providecommand \@@startlink[1]{}%
\providecommand \@@endlink[0]{}%
\providecommand \url  [0]{\begingroup\@sanitize@url \@url }%
\providecommand \@url [1]{\endgroup\@href {#1}{\urlprefix }}%
\providecommand \urlprefix  [0]{URL }%
\providecommand \Eprint [0]{\href }%
\providecommand \doibase [0]{http://dx.doi.org/}%
\providecommand \selectlanguage [0]{\@gobble}%
\providecommand \bibinfo  [0]{\@secondoftwo}%
\providecommand \bibfield  [0]{\@secondoftwo}%
\providecommand \translation [1]{[#1]}%
\providecommand \BibitemOpen [0]{}%
\providecommand \bibitemStop [0]{}%
\providecommand \bibitemNoStop [0]{.\EOS\space}%
\providecommand \EOS [0]{\spacefactor3000\relax}%
\providecommand \BibitemShut  [1]{\csname bibitem#1\endcsname}%
\let\auto@bib@innerbib\@empty
\bibitem [{\citenamefont {\mbox{v. Klitzing}}\ \emph
  {et~al.}(1980)\citenamefont {\mbox{v. Klitzing}}, \citenamefont {Dorda},\
  and\ \citenamefont {Pepper}}]{klitz80}%
  \BibitemOpen
  \bibfield  {author} {\bibinfo {author} {\bibfnamefont {K.}~\bibnamefont
  {\mbox{v. Klitzing}}}, \bibinfo {author} {\bibfnamefont {G.}~\bibnamefont
  {Dorda}}, \ and\ \bibinfo {author} {\bibfnamefont {M.}~\bibnamefont
  {Pepper}},\ }\href@noop {} {\bibfield  {journal} {\bibinfo  {journal} {Phys.
  Rev. Lett.}\ }\textbf {\bibinfo {volume} {45}},\ \bibinfo {pages} {494}
  (\bibinfo {year} {1980})}\BibitemShut {NoStop}%
\bibitem [{\citenamefont {Tsui}\ \emph {et~al.}(1982)\citenamefont {Tsui},
  \citenamefont {Stormer},\ and\ \citenamefont {Gossard}}]{tsui82}%
  \BibitemOpen
  \bibfield  {author} {\bibinfo {author} {\bibfnamefont {D.~C.}\ \bibnamefont
  {Tsui}}, \bibinfo {author} {\bibfnamefont {H.~L.}\ \bibnamefont {Stormer}}, \
  and\ \bibinfo {author} {\bibfnamefont {A.~C.}\ \bibnamefont {Gossard}},\
  }\href@noop {} {\bibfield  {journal} {\bibinfo  {journal} {Phys. Rev. Lett.}\
  }\textbf {\bibinfo {volume} {48}},\ \bibinfo {pages} {1559} (\bibinfo {year}
  {1982})}\BibitemShut {NoStop}%
\bibitem [{\citenamefont {Hatsugai}(1993)}]{hatsu93}%
  \BibitemOpen
  \bibfield  {author} {\bibinfo {author} {\bibfnamefont {Y.}~\bibnamefont
  {Hatsugai}},\ }\href@noop {} {\bibfield  {journal} {\bibinfo  {journal}
  {Phys. Rev. Lett.}\ }\textbf {\bibinfo {volume} {71}},\ \bibinfo {pages}
  {3697} (\bibinfo {year} {1993})}\BibitemShut {NoStop}%
\bibitem [{\citenamefont {Thouless}\ \emph {et~al.}(1982)\citenamefont
  {Thouless}, \citenamefont {Kohmoto}, \citenamefont {Nightingale},\ and\
  \citenamefont {den Nijs}}]{thoul82}%
  \BibitemOpen
  \bibfield  {author} {\bibinfo {author} {\bibfnamefont {D.~J.}\ \bibnamefont
  {Thouless}}, \bibinfo {author} {\bibfnamefont {M.}~\bibnamefont {Kohmoto}},
  \bibinfo {author} {\bibfnamefont {M.~P.}\ \bibnamefont {Nightingale}}, \ and\
  \bibinfo {author} {\bibfnamefont {M.}~\bibnamefont {den Nijs}},\ }\href@noop
  {} {\bibfield  {journal} {\bibinfo  {journal} {Phys. Rev. Lett.}\ }\textbf
  {\bibinfo {volume} {49}},\ \bibinfo {pages} {405} (\bibinfo {year}
  {1982})}\BibitemShut {NoStop}%
\bibitem [{\citenamefont {Avron}\ \emph {et~al.}(1983)\citenamefont {Avron},
  \citenamefont {Seiler},\ and\ \citenamefont {Simon}}]{avron83}%
  \BibitemOpen
  \bibfield  {author} {\bibinfo {author} {\bibfnamefont {J.~E.}\ \bibnamefont
  {Avron}}, \bibinfo {author} {\bibfnamefont {R.}~\bibnamefont {Seiler}}, \
  and\ \bibinfo {author} {\bibfnamefont {B.}~\bibnamefont {Simon}},\
  }\href@noop {} {\bibfield  {journal} {\bibinfo  {journal} {Phys. Rev. Lett.}\
  }\textbf {\bibinfo {volume} {51}},\ \bibinfo {pages} {51} (\bibinfo {year}
  {1983})}\BibitemShut {NoStop}%
\bibitem [{\citenamefont {Niu}\ \emph {et~al.}(1985)\citenamefont {Niu},
  \citenamefont {Thouless},\ and\ \citenamefont {Wu}}]{niu85}%
  \BibitemOpen
  \bibfield  {author} {\bibinfo {author} {\bibfnamefont {Q.}~\bibnamefont
  {Niu}}, \bibinfo {author} {\bibfnamefont {D.~J.}\ \bibnamefont {Thouless}}, \
  and\ \bibinfo {author} {\bibfnamefont {Y.-S.}\ \bibnamefont {Wu}},\
  }\href@noop {} {\bibfield  {journal} {\bibinfo  {journal} {Phys. Rev. B}\
  }\textbf {\bibinfo {volume} {31}},\ \bibinfo {pages} {3372} (\bibinfo {year}
  {1985})}\BibitemShut {NoStop}%
\bibitem [{\citenamefont {Kohmoto}(1985)}]{kohmo85}%
  \BibitemOpen
  \bibfield  {author} {\bibinfo {author} {\bibfnamefont {M.}~\bibnamefont
  {Kohmoto}},\ }\href@noop {} {\bibfield  {journal} {\bibinfo  {journal} {Ann.
  of Phys.}\ }\textbf {\bibinfo {volume} {160}},\ \bibinfo {pages} {343}
  (\bibinfo {year} {1985})}\BibitemShut {NoStop}%
\bibitem [{\citenamefont {Uhrig}(1991)}]{uhrig91}%
  \BibitemOpen
  \bibfield  {author} {\bibinfo {author} {\bibfnamefont {G.~S.}\ \bibnamefont
  {Uhrig}},\ }\href@noop {} {\bibfield  {journal} {\bibinfo  {journal} {Z.
  Phys. B}\ }\textbf {\bibinfo {volume} {82}},\ \bibinfo {pages} {29} (\bibinfo
  {year} {1991})}\BibitemShut {NoStop}%
\bibitem [{\citenamefont {Berry}(1984)}]{berry84}%
  \BibitemOpen
  \bibfield  {author} {\bibinfo {author} {\bibfnamefont {M.~V.}\ \bibnamefont
  {Berry}},\ }\href@noop {} {\bibfield  {journal} {\bibinfo  {journal} {Phys.
  Roy. Soc. Lond.}\ }\textbf {\bibinfo {volume} {A 392}},\ \bibinfo {pages}
  {45} (\bibinfo {year} {1984})}\BibitemShut {NoStop}%
\bibitem [{\citenamefont {Haldane}(1988)}]{halda88b}%
  \BibitemOpen
  \bibfield  {author} {\bibinfo {author} {\bibfnamefont {F.~D.~M.}\
  \bibnamefont {Haldane}},\ }\href@noop {} {\bibfield  {journal} {\bibinfo
  {journal} {Phys. Rev. Lett.}\ }\textbf {\bibinfo {volume} {61}},\ \bibinfo
  {pages} {2015} (\bibinfo {year} {1988})}\BibitemShut {NoStop}%
\bibitem [{\citenamefont {Redder}\ and\ \citenamefont {Uhrig}(2016)}]{redde16}%
  \BibitemOpen
  \bibfield  {author} {\bibinfo {author} {\bibfnamefont {C.~H.}\ \bibnamefont
  {Redder}}\ and\ \bibinfo {author} {\bibfnamefont {G.~S.}\ \bibnamefont
  {Uhrig}},\ }\href@noop {} {\bibfield  {journal} {\bibinfo  {journal} {Phys.
  Rev. A}\ }\textbf {\bibinfo {volume} {93}},\ \bibinfo {pages} {033654}
  (\bibinfo {year} {2016})}\BibitemShut {NoStop}%
\bibitem [{\citenamefont {Weng}\ \emph {et~al.}(2015)\citenamefont {Weng},
  \citenamefont {Yu}, \citenamefont {Hu}, \citenamefont {Dai},\ and\
  \citenamefont {Fang}}]{weng15}%
  \BibitemOpen
  \bibfield  {author} {\bibinfo {author} {\bibfnamefont {H.}~\bibnamefont
  {Weng}}, \bibinfo {author} {\bibfnamefont {R.}~\bibnamefont {Yu}}, \bibinfo
  {author} {\bibfnamefont {X.}~\bibnamefont {Hu}}, \bibinfo {author}
  {\bibfnamefont {X.}~\bibnamefont {Dai}}, \ and\ \bibinfo {author}
  {\bibfnamefont {Z.}~\bibnamefont {Fang}},\ }\href@noop {} {\bibfield
  {journal} {\bibinfo  {journal} {Adv. Phys.}\ }\textbf {\bibinfo {volume}
  {64}},\ \bibinfo {pages} {227} (\bibinfo {year} {2015})}\BibitemShut
  {NoStop}%
\bibitem [{\citenamefont {Liu}\ \emph {et~al.}(2016)\citenamefont {Liu},
  \citenamefont {Zhang},\ and\ \citenamefont {Qi}}]{liu16}%
  \BibitemOpen
  \bibfield  {author} {\bibinfo {author} {\bibfnamefont {C.-X.}\ \bibnamefont
  {Liu}}, \bibinfo {author} {\bibfnamefont {S.-C.}\ \bibnamefont {Zhang}}, \
  and\ \bibinfo {author} {\bibfnamefont {X.-L.}\ \bibnamefont {Qi}},\
  }\href@noop {} {\bibfield  {journal} {\bibinfo  {journal} {Annu. Rev.
  Condens. Matter Phys.}\ }\textbf {\bibinfo {volume} {7}},\ \bibinfo {pages}
  {301} (\bibinfo {year} {2016})}\BibitemShut {NoStop}%
\bibitem [{\citenamefont {Ren}\ \emph {et~al.}(2016)\citenamefont {Ren},
  \citenamefont {Qiao},\ and\ \citenamefont {Niu}}]{ren16}%
  \BibitemOpen
  \bibfield  {author} {\bibinfo {author} {\bibfnamefont {Y.}~\bibnamefont
  {Ren}}, \bibinfo {author} {\bibfnamefont {Z.}~\bibnamefont {Qiao}}, \ and\
  \bibinfo {author} {\bibfnamefont {Q.}~\bibnamefont {Niu}},\ }\href@noop {}
  {\bibfield  {journal} {\bibinfo  {journal} {Rep. Prog. Phys.}\ }\textbf
  {\bibinfo {volume} {79}},\ \bibinfo {pages} {066501} (\bibinfo {year}
  {2016})}\BibitemShut {NoStop}%
\bibitem [{\citenamefont {Kane}\ and\ \citenamefont
  {Mele}(2005{\natexlab{a}})}]{kane05a}%
  \BibitemOpen
  \bibfield  {author} {\bibinfo {author} {\bibfnamefont {C.~L.}\ \bibnamefont
  {Kane}}\ and\ \bibinfo {author} {\bibfnamefont {E.~J.}\ \bibnamefont
  {Mele}},\ }\href@noop {} {\bibfield  {journal} {\bibinfo  {journal} {Phys.
  Rev. Lett.}\ }\textbf {\bibinfo {volume} {95}},\ \bibinfo {pages} {146802}
  (\bibinfo {year} {2005}{\natexlab{a}})}\BibitemShut {NoStop}%
\bibitem [{\citenamefont {Kane}\ and\ \citenamefont
  {Mele}(2005{\natexlab{b}})}]{kane05b}%
  \BibitemOpen
  \bibfield  {author} {\bibinfo {author} {\bibfnamefont {C.~L.}\ \bibnamefont
  {Kane}}\ and\ \bibinfo {author} {\bibfnamefont {E.~J.}\ \bibnamefont
  {Mele}},\ }\href@noop {} {\bibfield  {journal} {\bibinfo  {journal} {Phys.
  Rev. Lett.}\ }\textbf {\bibinfo {volume} {95}},\ \bibinfo {pages} {226801}
  (\bibinfo {year} {2005}{\natexlab{b}})}\BibitemShut {NoStop}%
\bibitem [{\citenamefont {Hasan}\ and\ \citenamefont {Kane}(2010)}]{hasan10}%
  \BibitemOpen
  \bibfield  {author} {\bibinfo {author} {\bibfnamefont {M.~Z.}\ \bibnamefont
  {Hasan}}\ and\ \bibinfo {author} {\bibfnamefont {C.~L.}\ \bibnamefont
  {Kane}},\ }\href@noop {} {\bibfield  {journal} {\bibinfo  {journal} {Rev.
  Mod. Phys.}\ }\textbf {\bibinfo {volume} {82}},\ \bibinfo {pages} {3045}
  (\bibinfo {year} {2010})}\BibitemShut {NoStop}%
\bibitem [{\citenamefont {Fu}\ and\ \citenamefont {Kane}(2006)}]{fu06}%
  \BibitemOpen
  \bibfield  {author} {\bibinfo {author} {\bibfnamefont {L.}~\bibnamefont
  {Fu}}\ and\ \bibinfo {author} {\bibfnamefont {C.~L.}\ \bibnamefont {Kane}},\
  }\href@noop {} {\bibfield  {journal} {\bibinfo  {journal} {Phys. Rev. B}\
  }\textbf {\bibinfo {volume} {74}},\ \bibinfo {pages} {195312} (\bibinfo
  {year} {2006})}\BibitemShut {NoStop}%
\bibitem [{\citenamefont {Bernevig}\ and\ \citenamefont
  {Hughes}(2013)}]{berne13}%
  \BibitemOpen
  \bibfield  {author} {\bibinfo {author} {\bibfnamefont {A.~B.}\ \bibnamefont
  {Bernevig}}\ and\ \bibinfo {author} {\bibfnamefont {T.~L.}\ \bibnamefont
  {Hughes}},\ }\href@noop {} {\emph {\bibinfo {title} {Topological Insulators
  and Topological Superconductors}}}\ (\bibinfo  {publisher} {Princeton
  University Press},\ \bibinfo {address} {Princeton},\ \bibinfo {year}
  {2013})\BibitemShut {NoStop}%
\bibitem [{\citenamefont {Wu}\ \emph {et~al.}(2006)\citenamefont {Wu},
  \citenamefont {Bernevig},\ and\ \citenamefont {Zhang}}]{wu06}%
  \BibitemOpen
  \bibfield  {author} {\bibinfo {author} {\bibfnamefont {C.}~\bibnamefont
  {Wu}}, \bibinfo {author} {\bibfnamefont {B.~A.}\ \bibnamefont {Bernevig}}, \
  and\ \bibinfo {author} {\bibfnamefont {S.-C.}\ \bibnamefont {Zhang}},\
  }\href@noop {} {\bibfield  {journal} {\bibinfo  {journal} {Phys. Rev. Lett.}\
  }\textbf {\bibinfo {volume} {96}},\ \bibinfo {pages} {106401} (\bibinfo
  {year} {2006})}\BibitemShut {NoStop}%
\bibitem [{\citenamefont {Qi}\ and\ \citenamefont {Zhang}(2011)}]{qi11}%
  \BibitemOpen
  \bibfield  {author} {\bibinfo {author} {\bibfnamefont {X.-L.}\ \bibnamefont
  {Qi}}\ and\ \bibinfo {author} {\bibfnamefont {S.-C.}\ \bibnamefont {Zhang}},\
  }\href@noop {} {\bibfield  {journal} {\bibinfo  {journal} {Rev. Mod. Phys.}\
  }\textbf {\bibinfo {volume} {83}},\ \bibinfo {pages} {1058} (\bibinfo {year}
  {2011})}\BibitemShut {NoStop}%
\bibitem [{\citenamefont {Bernevig}\ \emph {et~al.}(2006)\citenamefont
  {Bernevig}, \citenamefont {Hughes},\ and\ \citenamefont {Zhang}}]{berne06}%
  \BibitemOpen
  \bibfield  {author} {\bibinfo {author} {\bibfnamefont {B.~A.}\ \bibnamefont
  {Bernevig}}, \bibinfo {author} {\bibfnamefont {T.~L.}\ \bibnamefont
  {Hughes}}, \ and\ \bibinfo {author} {\bibfnamefont {S.-C.}\ \bibnamefont
  {Zhang}},\ }\href@noop {} {\bibfield  {journal} {\bibinfo  {journal}
  {Science}\ }\textbf {\bibinfo {volume} {314}},\ \bibinfo {pages} {1757}
  (\bibinfo {year} {2006})}\BibitemShut {NoStop}%
\bibitem [{\citenamefont {K\"onig}\ \emph {et~al.}(2007)\citenamefont
  {K\"onig}, \citenamefont {Wiedmann}, \citenamefont {Br\"une}, \citenamefont
  {Roth}, \citenamefont {Buhmann}, \citenamefont {Molenkamp}, \citenamefont
  {Qi},\ and\ \citenamefont {Zhang}}]{konig07}%
  \BibitemOpen
  \bibfield  {author} {\bibinfo {author} {\bibfnamefont {M.}~\bibnamefont
  {K\"onig}}, \bibinfo {author} {\bibfnamefont {S.}~\bibnamefont {Wiedmann}},
  \bibinfo {author} {\bibfnamefont {C.}~\bibnamefont {Br\"une}}, \bibinfo
  {author} {\bibfnamefont {A.}~\bibnamefont {Roth}}, \bibinfo {author}
  {\bibfnamefont {H.}~\bibnamefont {Buhmann}}, \bibinfo {author} {\bibfnamefont
  {L.~W.}\ \bibnamefont {Molenkamp}}, \bibinfo {author} {\bibfnamefont {X.-L.}\
  \bibnamefont {Qi}}, \ and\ \bibinfo {author} {\bibfnamefont {S.-C.}\
  \bibnamefont {Zhang}},\ }\href@noop {} {\bibfield  {journal} {\bibinfo
  {journal} {Science}\ }\textbf {\bibinfo {volume} {318}},\ \bibinfo {pages}
  {766} (\bibinfo {year} {2007})}\BibitemShut {NoStop}%
\bibitem [{\citenamefont {Roth}\ \emph {et~al.}(2009)\citenamefont {Roth},
  \citenamefont {Br\"une}, \citenamefont {Buhmann}, \citenamefont {Molenkamp},
  \citenamefont {Maciejko}, \citenamefont {Qi},\ and\ \citenamefont
  {Zhang}}]{roth09}%
  \BibitemOpen
  \bibfield  {author} {\bibinfo {author} {\bibfnamefont {A.}~\bibnamefont
  {Roth}}, \bibinfo {author} {\bibfnamefont {C.}~\bibnamefont {Br\"une}},
  \bibinfo {author} {\bibfnamefont {H.}~\bibnamefont {Buhmann}}, \bibinfo
  {author} {\bibfnamefont {L.~W.}\ \bibnamefont {Molenkamp}}, \bibinfo {author}
  {\bibfnamefont {J.}~\bibnamefont {Maciejko}}, \bibinfo {author}
  {\bibfnamefont {X.-L.}\ \bibnamefont {Qi}}, \ and\ \bibinfo {author}
  {\bibfnamefont {S.-C.}\ \bibnamefont {Zhang}},\ }\href@noop {} {\bibfield
  {journal} {\bibinfo  {journal} {Science}\ }\textbf {\bibinfo {volume}
  {325}},\ \bibinfo {pages} {294} (\bibinfo {year} {2009})}\BibitemShut
  {NoStop}%
\bibitem [{\citenamefont {Knez}\ \emph {et~al.}(2011)\citenamefont {Knez},
  \citenamefont {Du},\ and\ \citenamefont {Sullivan}}]{knez11}%
  \BibitemOpen
  \bibfield  {author} {\bibinfo {author} {\bibfnamefont {I.}~\bibnamefont
  {Knez}}, \bibinfo {author} {\bibfnamefont {R.-R.}\ \bibnamefont {Du}}, \ and\
  \bibinfo {author} {\bibfnamefont {G.}~\bibnamefont {Sullivan}},\ }\href@noop
  {} {\bibfield  {journal} {\bibinfo  {journal} {Phys. Rev. Lett.}\ }\textbf
  {\bibinfo {volume} {107}},\ \bibinfo {pages} {136603} (\bibinfo {year}
  {2011})}\BibitemShut {NoStop}%
\bibitem [{\citenamefont {Du}\ \emph {et~al.}(2015)\citenamefont {Du},
  \citenamefont {Knez}, \citenamefont {Sullivan},\ and\ \citenamefont
  {Du}}]{du15}%
  \BibitemOpen
  \bibfield  {author} {\bibinfo {author} {\bibfnamefont {L.}~\bibnamefont
  {Du}}, \bibinfo {author} {\bibfnamefont {I.}~\bibnamefont {Knez}}, \bibinfo
  {author} {\bibfnamefont {G.}~\bibnamefont {Sullivan}}, \ and\ \bibinfo
  {author} {\bibfnamefont {R.-R.}\ \bibnamefont {Du}},\ }\href@noop {}
  {\bibfield  {journal} {\bibinfo  {journal} {Phys. Rev. Lett.}\ }\textbf
  {\bibinfo {volume} {114}},\ \bibinfo {pages} {096802} (\bibinfo {year}
  {2015})}\BibitemShut {NoStop}%
\bibitem [{\citenamefont {Liu}\ \emph {et~al.}(2011)\citenamefont {Liu},
  \citenamefont {Feng},\ and\ \citenamefont {Yao}}]{liu11}%
  \BibitemOpen
  \bibfield  {author} {\bibinfo {author} {\bibfnamefont {C.-C.}\ \bibnamefont
  {Liu}}, \bibinfo {author} {\bibfnamefont {W.}~\bibnamefont {Feng}}, \ and\
  \bibinfo {author} {\bibfnamefont {Y.}~\bibnamefont {Yao}},\ }\href@noop {}
  {\bibfield  {journal} {\bibinfo  {journal} {Phys. Rev. Lett.}\ }\textbf
  {\bibinfo {volume} {107}},\ \bibinfo {pages} {076802} (\bibinfo {year}
  {2011})}\BibitemShut {NoStop}%
\bibitem [{\citenamefont {Chang}\ \emph {et~al.}(2013)\citenamefont {Chang},
  \citenamefont {Zhang}, \citenamefont {Feng}, \citenamefont {Shen},
  \citenamefont {Zhang}, \citenamefont {Guo}, \citenamefont {Li}, \citenamefont
  {Ou}, \citenamefont {Wei}, \citenamefont {Wang}, \citenamefont {Ji},
  \citenamefont {Feng}, \citenamefont {Ji}, \citenamefont {Chen}, \citenamefont
  {Jia}, \citenamefont {Dai}, \citenamefont {Fang}, \citenamefont {Zhang},
  \citenamefont {He}, \citenamefont {Wang}, \citenamefont {Lu}, \citenamefont
  {Ma},\ and\ \citenamefont {Xue}}]{chang13}%
  \BibitemOpen
  \bibfield  {author} {\bibinfo {author} {\bibfnamefont {C.-Z.}\ \bibnamefont
  {Chang}}, \bibinfo {author} {\bibfnamefont {J.}~\bibnamefont {Zhang}},
  \bibinfo {author} {\bibfnamefont {X.}~\bibnamefont {Feng}}, \bibinfo {author}
  {\bibfnamefont {J.}~\bibnamefont {Shen}}, \bibinfo {author} {\bibfnamefont
  {Z.}~\bibnamefont {Zhang}}, \bibinfo {author} {\bibfnamefont
  {M.}~\bibnamefont {Guo}}, \bibinfo {author} {\bibfnamefont {K.}~\bibnamefont
  {Li}}, \bibinfo {author} {\bibfnamefont {Y.}~\bibnamefont {Ou}}, \bibinfo
  {author} {\bibfnamefont {P.}~\bibnamefont {Wei}}, \bibinfo {author}
  {\bibfnamefont {L.-L.}\ \bibnamefont {Wang}}, \bibinfo {author}
  {\bibfnamefont {Z.-Q.}\ \bibnamefont {Ji}}, \bibinfo {author} {\bibfnamefont
  {Y.}~\bibnamefont {Feng}}, \bibinfo {author} {\bibfnamefont {S.}~\bibnamefont
  {Ji}}, \bibinfo {author} {\bibfnamefont {X.}~\bibnamefont {Chen}}, \bibinfo
  {author} {\bibfnamefont {J.}~\bibnamefont {Jia}}, \bibinfo {author}
  {\bibfnamefont {X.}~\bibnamefont {Dai}}, \bibinfo {author} {\bibfnamefont
  {Z.}~\bibnamefont {Fang}}, \bibinfo {author} {\bibfnamefont {S.-C.}\
  \bibnamefont {Zhang}}, \bibinfo {author} {\bibfnamefont {K.}~\bibnamefont
  {He}}, \bibinfo {author} {\bibfnamefont {Y.}~\bibnamefont {Wang}}, \bibinfo
  {author} {\bibfnamefont {L.}~\bibnamefont {Lu}}, \bibinfo {author}
  {\bibfnamefont {X.-C.}\ \bibnamefont {Ma}}, \ and\ \bibinfo {author}
  {\bibfnamefont {Q.-K.}\ \bibnamefont {Xue}},\ }\href@noop {} {\bibfield
  {journal} {\bibinfo  {journal} {Science}\ }\textbf {\bibinfo {volume}
  {340}},\ \bibinfo {pages} {167} (\bibinfo {year} {2013})}\BibitemShut
  {NoStop}%
\bibitem [{\citenamefont {Kou}\ \emph {et~al.}(2014)\citenamefont {Kou},
  \citenamefont {Guo}, \citenamefont {Fan}, \citenamefont {Pan}, \citenamefont
  {Lang}, \citenamefont {Jiang}, \citenamefont {Shao}, \citenamefont {Nie},
  \citenamefont {Murata}, \citenamefont {Tang}, \citenamefont {Wang},
  \citenamefont {He}, \citenamefont {Lee}, \citenamefont {Lee},\ and\
  \citenamefont {Wang}}]{kou14}%
  \BibitemOpen
  \bibfield  {author} {\bibinfo {author} {\bibfnamefont {X.}~\bibnamefont
  {Kou}}, \bibinfo {author} {\bibfnamefont {S.-T.}\ \bibnamefont {Guo}},
  \bibinfo {author} {\bibfnamefont {Y.}~\bibnamefont {Fan}}, \bibinfo {author}
  {\bibfnamefont {L.}~\bibnamefont {Pan}}, \bibinfo {author} {\bibfnamefont
  {M.}~\bibnamefont {Lang}}, \bibinfo {author} {\bibfnamefont {Y.}~\bibnamefont
  {Jiang}}, \bibinfo {author} {\bibfnamefont {Q.}~\bibnamefont {Shao}},
  \bibinfo {author} {\bibfnamefont {T.}~\bibnamefont {Nie}}, \bibinfo {author}
  {\bibfnamefont {K.}~\bibnamefont {Murata}}, \bibinfo {author} {\bibfnamefont
  {J.}~\bibnamefont {Tang}}, \bibinfo {author} {\bibfnamefont {Y.}~\bibnamefont
  {Wang}}, \bibinfo {author} {\bibfnamefont {L.}~\bibnamefont {He}}, \bibinfo
  {author} {\bibfnamefont {T.-K.}\ \bibnamefont {Lee}}, \bibinfo {author}
  {\bibfnamefont {W.-L.}\ \bibnamefont {Lee}}, \ and\ \bibinfo {author}
  {\bibfnamefont {K.~L.}\ \bibnamefont {Wang}},\ }\href@noop {} {\bibfield
  {journal} {\bibinfo  {journal} {Phys. Rev. Lett.}\ }\textbf {\bibinfo
  {volume} {113}},\ \bibinfo {pages} {137201} (\bibinfo {year}
  {2014})}\BibitemShut {NoStop}%
\bibitem [{\citenamefont {Chang}\ \emph {et~al.}(2015)\citenamefont {Chang},
  \citenamefont {Zhao}, \citenamefont {Kim}, \citenamefont {Zhang},
  \citenamefont {Assaf}, \citenamefont {Heiman}, \citenamefont {Zhang},
  \citenamefont {Liu}, \citenamefont {Chan},\ and\ \citenamefont
  {Moodera}}]{chang15}%
  \BibitemOpen
  \bibfield  {author} {\bibinfo {author} {\bibfnamefont {C.-Z.}\ \bibnamefont
  {Chang}}, \bibinfo {author} {\bibfnamefont {W.}~\bibnamefont {Zhao}},
  \bibinfo {author} {\bibfnamefont {D.~Y.}\ \bibnamefont {Kim}}, \bibinfo
  {author} {\bibfnamefont {H.}~\bibnamefont {Zhang}}, \bibinfo {author}
  {\bibfnamefont {B.~A.}\ \bibnamefont {Assaf}}, \bibinfo {author}
  {\bibfnamefont {D.}~\bibnamefont {Heiman}}, \bibinfo {author} {\bibfnamefont
  {S.-C.}\ \bibnamefont {Zhang}}, \bibinfo {author} {\bibfnamefont
  {C.}~\bibnamefont {Liu}}, \bibinfo {author} {\bibfnamefont {M.~H.~W.}\
  \bibnamefont {Chan}}, \ and\ \bibinfo {author} {\bibfnamefont {J.~S.}\
  \bibnamefont {Moodera}},\ }\href@noop {} {\bibfield  {journal} {\bibinfo
  {journal} {Nature Mat.}\ }\textbf {\bibinfo {volume} {14}},\ \bibinfo {pages}
  {473} (\bibinfo {year} {2015})}\BibitemShut {NoStop}%
\bibitem [{\citenamefont {Wu}\ \emph {et~al.}(2014)\citenamefont {Wu},
  \citenamefont {Shan},\ and\ \citenamefont {Yan}}]{wu14}%
  \BibitemOpen
  \bibfield  {author} {\bibinfo {author} {\bibfnamefont {S.-C.}\ \bibnamefont
  {Wu}}, \bibinfo {author} {\bibfnamefont {G.}~\bibnamefont {Shan}}, \ and\
  \bibinfo {author} {\bibfnamefont {B.}~\bibnamefont {Yan}},\ }\href@noop {}
  {\bibfield  {journal} {\bibinfo  {journal} {Phys. Rev. Lett.}\ }\textbf
  {\bibinfo {volume} {113}},\ \bibinfo {pages} {256401} (\bibinfo {year}
  {2014})}\BibitemShut {NoStop}%
\bibitem [{\citenamefont {Han}\ \emph {et~al.}(2015)\citenamefont {Han},
  \citenamefont {Wan}, \citenamefont {Ge}, \citenamefont {Song},\ and\
  \citenamefont {Wang}}]{han15}%
  \BibitemOpen
  \bibfield  {author} {\bibinfo {author} {\bibfnamefont {Y.}~\bibnamefont
  {Han}}, \bibinfo {author} {\bibfnamefont {J.-G.}\ \bibnamefont {Wan}},
  \bibinfo {author} {\bibfnamefont {G.-X.}\ \bibnamefont {Ge}}, \bibinfo
  {author} {\bibfnamefont {F.-Q.}\ \bibnamefont {Song}}, \ and\ \bibinfo
  {author} {\bibfnamefont {G.-H.}\ \bibnamefont {Wang}},\ }\href@noop {}
  {\bibfield  {journal} {\bibinfo  {journal} {Sci. Rep.}\ }\textbf {\bibinfo
  {volume} {5}},\ \bibinfo {pages} {16843} (\bibinfo {year}
  {2015})}\BibitemShut {NoStop}%
\bibitem [{\citenamefont {Krasheninnikov}\ and\ \citenamefont
  {Nieminen}(2011)}]{krash11}%
  \BibitemOpen
  \bibfield  {author} {\bibinfo {author} {\bibfnamefont {A.~V.}\ \bibnamefont
  {Krasheninnikov}}\ and\ \bibinfo {author} {\bibfnamefont {R.~M.}\
  \bibnamefont {Nieminen}},\ }\href@noop {} {\bibfield  {journal} {\bibinfo
  {journal} {Theor. Chem. Acc}\ }\textbf {\bibinfo {volume} {129}},\ \bibinfo
  {pages} {625} (\bibinfo {year} {2011})}\BibitemShut {NoStop}%
\bibitem [{\citenamefont {Jotzu}\ \emph {et~al.}(2014)\citenamefont {Jotzu},
  \citenamefont {Messer}, \citenamefont {Desbuquois}, \citenamefont {Lebrat},
  \citenamefont {Uehlinger}, \citenamefont {Greif},\ and\ \citenamefont
  {Esslinger}}]{jotzu14}%
  \BibitemOpen
  \bibfield  {author} {\bibinfo {author} {\bibfnamefont {G.}~\bibnamefont
  {Jotzu}}, \bibinfo {author} {\bibfnamefont {M.}~\bibnamefont {Messer}},
  \bibinfo {author} {\bibfnamefont {R.}~\bibnamefont {Desbuquois}}, \bibinfo
  {author} {\bibfnamefont {M.}~\bibnamefont {Lebrat}}, \bibinfo {author}
  {\bibfnamefont {T.}~\bibnamefont {Uehlinger}}, \bibinfo {author}
  {\bibfnamefont {D.}~\bibnamefont {Greif}}, \ and\ \bibinfo {author}
  {\bibfnamefont {T.}~\bibnamefont {Esslinger}},\ }\href@noop {} {\bibfield
  {journal} {\bibinfo  {journal} {Nature}\ }\textbf {\bibinfo {volume} {515}},\
  \bibinfo {pages} {237} (\bibinfo {year} {2014})}\BibitemShut {NoStop}%
\bibitem [{\citenamefont {Van Dyke}\ and\ \citenamefont {Morr}(2016)}]{dyke16}%
  \BibitemOpen
  \bibfield  {author} {\bibinfo {author} {\bibfnamefont {J.~S.}\
  \bibnamefont {Van Dyke}}\ and\ \bibinfo {author} {\bibfnamefont {D.~K.}\
  \bibnamefont {Morr}},\ }\href@noop {} {\bibfield  {journal} {\bibinfo
  {journal} {Phys. Rev. B}\ }\textbf {\bibinfo {volume} {93}},\ \bibinfo
  {pages} {081401(R)} (\bibinfo {year} {2016})}\BibitemShut {NoStop}%
\bibitem [{\citenamefont {Qiao}\ \emph {et~al.}(2016)\citenamefont {Qiao},
  \citenamefont {Han}, \citenamefont {Zhang}, \citenamefont {Wang},
  \citenamefont {Deng}, \citenamefont {Jiang}, \citenamefont {Yang},
  \citenamefont {Wang},\ and\ \citenamefont {Niu}}]{qiao16}%
  \BibitemOpen
  \bibfield  {author} {\bibinfo {author} {\bibfnamefont {Z.}~\bibnamefont
  {Qiao}}, \bibinfo {author} {\bibfnamefont {Y.}~\bibnamefont {Han}}, \bibinfo
  {author} {\bibfnamefont {L.}~\bibnamefont {Zhang}}, \bibinfo {author}
  {\bibfnamefont {K.}~\bibnamefont {Wang}}, \bibinfo {author} {\bibfnamefont
  {X.}~\bibnamefont {Deng}}, \bibinfo {author} {\bibfnamefont {H.}~\bibnamefont
  {Jiang}}, \bibinfo {author} {\bibfnamefont {S.~A.}\ \bibnamefont {Yang}},
  \bibinfo {author} {\bibfnamefont {J.}~\bibnamefont {Wang}}, \ and\ \bibinfo
  {author} {\bibfnamefont {Q.}~\bibnamefont {Niu}},\ }\href@noop {} {\bibfield
  {journal} {\bibinfo  {journal} {Phys. Rev. Lett.}\ }\textbf {\bibinfo
  {volume} {117}},\ \bibinfo {pages} {056802} (\bibinfo {year}
  {2016})}\BibitemShut {NoStop}%
\bibitem [{\citenamefont {Uhrig}(2016)}]{uhrig16}%
  \BibitemOpen
  \bibfield  {author} {\bibinfo {author} {\bibfnamefont {G.~S.}\ \bibnamefont
  {Uhrig}},\ }\href@noop {} {\bibfield  {journal} {\bibinfo  {journal} {Phys.
  Rev. B}\ }\textbf {\bibinfo {volume} {93}},\ \bibinfo {pages} {205438}
  (\bibinfo {year} {2016})}\BibitemShut {NoStop}%
\bibitem [{\citenamefont {Bychkov}\ and\ \citenamefont
  {Rashba}(1984)}]{bychk84}%
  \BibitemOpen
  \bibfield  {author} {\bibinfo {author} {\bibfnamefont {Y.~A.}\ \bibnamefont
  {Bychkov}}\ and\ \bibinfo {author} {\bibfnamefont {E.~I.}\ \bibnamefont
  {Rashba}},\ }\href@noop {} {\bibfield  {journal} {\bibinfo  {journal} {J.
  Phys. C}\ }\textbf {\bibinfo {volume} {17}},\ \bibinfo {pages} {6039}
  (\bibinfo {year} {1984})}\BibitemShut {NoStop}%
\bibitem [{\citenamefont {Wolf}\ \emph {et~al.}(2001)\citenamefont {Wolf},
  \citenamefont {Awschalom}, \citenamefont {Buhrman}, \citenamefont {Daughton},
  \citenamefont {von Moln\'ar}, \citenamefont {Roukes}, \citenamefont
  {Chtchelkanova},\ and\ \citenamefont {Treger}}]{wolf01}%
  \BibitemOpen
  \bibfield  {author} {\bibinfo {author} {\bibfnamefont {S.~A.}\ \bibnamefont
  {Wolf}}, \bibinfo {author} {\bibfnamefont {D.~D.}\ \bibnamefont {Awschalom}},
  \bibinfo {author} {\bibfnamefont {R.~A.}\ \bibnamefont {Buhrman}}, \bibinfo
  {author} {\bibfnamefont {J.~M.}\ \bibnamefont {Daughton}}, \bibinfo {author}
  {\bibfnamefont {S.}~\bibnamefont {von Moln\'ar}}, \bibinfo {author}
  {\bibfnamefont {M.~L.}\ \bibnamefont {Roukes}}, \bibinfo {author}
  {\bibfnamefont {A.~Y.}\ \bibnamefont {Chtchelkanova}}, \ and\ \bibinfo
  {author} {\bibfnamefont {D.~M.}\ \bibnamefont {Treger}},\ }\href@noop {}
  {\bibfield  {journal} {\bibinfo  {journal} {Science}\ }\textbf {\bibinfo
  {volume} {294}},\ \bibinfo {pages} {1488} (\bibinfo {year}
  {2001})}\BibitemShut {NoStop}%
\bibitem [{\citenamefont {\v{Z}uti\'c}\ \emph {et~al.}(2004)\citenamefont
  {\v{Z}uti\'c}, \citenamefont {Fabian},\ and\ \citenamefont {\mbox{Das
  Sarma}}}]{zutic04}%
  \BibitemOpen
  \bibfield  {author} {\bibinfo {author} {\bibfnamefont {I.}~\bibnamefont
  {\v{Z}uti\'c}}, \bibinfo {author} {\bibfnamefont {J.}~\bibnamefont {Fabian}},
  \ and\ \bibinfo {author} {\bibfnamefont {S.}~\bibnamefont {\mbox{Das
  Sarma}}},\ }\href@noop {} {\bibfield  {journal} {\bibinfo  {journal} {Rev.
  Mod. Phys.}\ }\textbf {\bibinfo {volume} {76}},\ \bibinfo {pages} {323}
  (\bibinfo {year} {2004})}\BibitemShut {NoStop}%
\bibitem [{\citenamefont {Kramers}(1930)}]{krame30}%
  \BibitemOpen
  \bibfield  {author} {\bibinfo {author} {\bibfnamefont {H.~A.}\ \bibnamefont
  {Kramers}},\ }\href@noop {} {\bibfield  {journal} {\bibinfo  {journal} {Proc.
  Koninklijke Akademie van Wetenschappen}\ }\textbf {\bibinfo {volume} {33}},\
  \bibinfo {pages} {959} (\bibinfo {year} {1930})}\BibitemShut {NoStop}%
\bibitem [{\citenamefont {Jiang}\ \emph {et~al.}(2016)\citenamefont {Jiang},
  \citenamefont {Chang}, \citenamefont {Tang}, \citenamefont {Zheng},
  \citenamefont {Moodera},\ and\ \citenamefont {Shi}}]{jiang16}%
  \BibitemOpen
  \bibfield  {author} {\bibinfo {author} {\bibfnamefont {Z.}~\bibnamefont
  {Jiang}}, \bibinfo {author} {\bibfnamefont {C.-Z.}\ \bibnamefont {Chang}},
  \bibinfo {author} {\bibfnamefont {C.}~\bibnamefont {Tang}}, \bibinfo {author}
  {\bibfnamefont {J.-G.}\ \bibnamefont {Zheng}}, \bibinfo {author}
  {\bibfnamefont {J.~S.}\ \bibnamefont {Moodera}}, \ and\ \bibinfo {author}
  {\bibfnamefont {J.}~\bibnamefont {Shi}},\ }\href@noop {} {\bibfield
  {journal} {\bibinfo  {journal} {AIP Advances}\ }\textbf {\bibinfo {volume}
  {6}},\ \bibinfo {pages} {055809} (\bibinfo {year} {2016})}\BibitemShut
  {NoStop}%
\bibitem [{\citenamefont {Jungwirth}\ \emph {et~al.}(2006)\citenamefont
  {Jungwirth}, \citenamefont {Sinova}, \citenamefont {Ma\v{s}ek}, \citenamefont
  {Ku\v{c}era},\ and\ \citenamefont {MacDonald}}]{jungw06}%
  \BibitemOpen
  \bibfield  {author} {\bibinfo {author} {\bibfnamefont {T.}~\bibnamefont
  {Jungwirth}}, \bibinfo {author} {\bibfnamefont {J.}~\bibnamefont {Sinova}},
  \bibinfo {author} {\bibfnamefont {J.}~\bibnamefont {Ma\v{s}ek}}, \bibinfo
  {author} {\bibfnamefont {J.}~\bibnamefont {Ku\v{c}era}}, \ and\ \bibinfo
  {author} {\bibfnamefont {A.~H.}\ \bibnamefont {MacDonald}},\ }\href@noop {}
  {\bibfield  {journal} {\bibinfo  {journal} {Rev. Mod. Phys.}\ }\textbf
  {\bibinfo {volume} {78}},\ \bibinfo {pages} {809} (\bibinfo {year}
  {2006})}\BibitemShut {NoStop}%
\bibitem [{\citenamefont {Qiao}\ \emph {et~al.}(2010)\citenamefont {Qiao},
  \citenamefont {Yang}, \citenamefont {Feng}, \citenamefont {Tse},
  \citenamefont {Ding}, \citenamefont {Yao}, \citenamefont {Wang},\ and\
  \citenamefont {Niu}}]{qiao10}%
  \BibitemOpen
  \bibfield  {author} {\bibinfo {author} {\bibfnamefont {Z.}~\bibnamefont
  {Qiao}}, \bibinfo {author} {\bibfnamefont {S.~A.}\ \bibnamefont {Yang}},
  \bibinfo {author} {\bibfnamefont {W.}~\bibnamefont {Feng}}, \bibinfo {author}
  {\bibfnamefont {W.-K.}\ \bibnamefont {Tse}}, \bibinfo {author} {\bibfnamefont
  {J.}~\bibnamefont {Ding}}, \bibinfo {author} {\bibfnamefont {Y.}~\bibnamefont
  {Yao}}, \bibinfo {author} {\bibfnamefont {J.}~\bibnamefont {Wang}}, \ and\
  \bibinfo {author} {\bibfnamefont {Q.}~\bibnamefont {Niu}},\ }\href@noop {}
  {\bibfield  {journal} {\bibinfo  {journal} {Phys. Rev. B}\ }\textbf {\bibinfo
  {volume} {82}},\ \bibinfo {pages} {161414(R)} (\bibinfo {year}
  {2010})}\BibitemShut {NoStop}%
\bibitem [{\citenamefont {Chen}\ \emph {et~al.}(2011)\citenamefont {Chen},
  \citenamefont {Xiao}, \citenamefont {Chiou},\ and\ \citenamefont
  {Guo}}]{chen11b}%
  \BibitemOpen
  \bibfield  {author} {\bibinfo {author} {\bibfnamefont {T.-W.}\ \bibnamefont
  {Chen}}, \bibinfo {author} {\bibfnamefont {Z.-R.}\ \bibnamefont {Xiao}},
  \bibinfo {author} {\bibfnamefont {D.-W.}\ \bibnamefont {Chiou}}, \ and\
  \bibinfo {author} {\bibfnamefont {G.-Y.}\ \bibnamefont {Guo}},\ }\href@noop
  {} {\bibfield  {journal} {\bibinfo  {journal} {Phys. Rev. B}\ }\textbf
  {\bibinfo {volume} {84}},\ \bibinfo {pages} {165453} (\bibinfo {year}
  {2011})}\BibitemShut {NoStop}%
\bibitem [{\citenamefont {Fruchart}\ and\ \citenamefont
  {Carpentier}(2013)}]{fruch13}%
  \BibitemOpen
  \bibfield  {author} {\bibinfo {author} {\bibfnamefont {M.}~\bibnamefont
  {Fruchart}}\ and\ \bibinfo {author} {\bibfnamefont {D.}~\bibnamefont
  {Carpentier}},\ }\href@noop {} {\bibfield  {journal} {\bibinfo  {journal}
  {Comptes Rendus Physique}\ }\textbf {\bibinfo {volume} {14}},\ \bibinfo
  {pages} {779} (\bibinfo {year} {2013})}\BibitemShut {NoStop}%
\bibitem [{\citenamefont {Bauer}\ and\ \citenamefont
  {Bobisch}(2016)}]{bauer16}%
  \BibitemOpen
  \bibfield  {author} {\bibinfo {author} {\bibfnamefont {S.}~\bibnamefont
  {Bauer}}\ and\ \bibinfo {author} {\bibfnamefont {C.~A.}\ \bibnamefont
  {Bobisch}},\ }\href@noop {} {\bibfield  {journal} {\bibinfo  {journal}
  {Nature Comm.}\ }\textbf {\bibinfo {volume} {7}},\ \bibinfo {pages} {11381}
  (\bibinfo {year} {2016})}\BibitemShut {NoStop}%
\end{thebibliography}

%

\end{document}